\documentclass[letterpaper,11pt,fleqn]{article}
\pdfoutput=1
\usepackage{jheppub}

\usepackage{graphicx}
\usepackage[figuresright]{rotating}
\usepackage{bm,amsmath,amssymb}
\usepackage[mathscr]{eucal}
\usepackage{color}

\long\def\comment#1{ }
\newcommand{\eqn}[1]{Eq.~\eqref{#1}}
\newcommand{\beq}{\begin{equation}}
\newcommand{\eeq}{\end{equation}}
\newcommand{\nn}{\nonumber\\}

\newcommand{\rmd}{{\rm d}}
\newcommand{\rme}{{\rm e}}

\newcommand{\del}{\partial}

\newcommand{\order}[1]{\mcal{O}{\left(#1\right)}}
\newcommand{\mcal}{\mathcal}

\newcommand{\abar}{\bar{\alpha}_s}

\newcommand{\tbr}{t_{\rm br}}

\newcommand{\obr}{\omega_{\rm br}}

\title{\Large Multi-particle correlations and
KNO scaling in the medium-induced jet evolution} 

\author{Miguel A.~Escobedo}

\author{and Edmond Iancu}

\affiliation{Institut de physique th\'{e}orique, Universit\'{e} Paris Saclay, CNRS, CEA, F-91191 Gif-sur-Yvette, France}

\emailAdd{Miguel-Angel.Escobedo-Espinosa@cea.fr}
\emailAdd{Edmond.Iancu@cea.fr}

\abstract{We study the gluon distribution produced via successive medium-induced branchings
by an energetic jet propagating through a weakly-coupled quark-gluon plasma. We show that
under suitable approximations the evolution of the jet can be described as a
classical stochastic process, which is exactly solvable. For this process, we construct
exact analytic solutions for all the $n$-point correlation functions (the $n$-body densities
in the space of energy). The corresponding results for the one-point and the two-point functions
were already known, but those for the higher-point functions are new. These results demonstrate
strong correlations associated with the existence of common ancestors in the branching process.
By integrating these $n$-point functions over the gluon energies, we deduce the mean gluon 
multiplicity $\langle N\rangle$ as well as the higher moments $\langle N^p\rangle$ with $p\ge 2$. 
We find that the multiplicities of the soft gluons are parametrically large and show a remarkable regularity, 
known as Koba-Nielsen-Olesen (KNO) scaling: the reduced moments 
$\langle N^p\rangle/\langle N\rangle^p$ are pure numbers, independent of any of the physical
parameters of the problem. We recognize a special negative binomial distribution which
is characterized by large statistical fluctuations. These predictions can be tested
in Pb+Pb collisions at the LHC, via event-by-event measurements of the di-jet asymmetry.
}

\keywords{Perturbative QCD. Heavy Ion Collisions. Di-jet asymmetry. Multi-particle correlations}

\begin{document}
\maketitle

\section{Introduction}
\label{sec:intro}

Motivated by extensive experimental studies of the energy loss by jets or leading
hadrons in ultrarelativistic nucleus-nucleus collisions at RHIC and the LHC, and notably by the
remarkable phenomenon known as `di-jet asymmetry' 
 \cite{Aad:2010bu,Chatrchyan:2011sx,Chatrchyan:2012nia,Aad:2012vca,Chatrchyan:2013kwa,Chatrchyan:2014ava,Aad:2014wha,Khachatryan:2015lha,Khachatryan:2016erx}, 
there has been a renewal of the interest in the theory and phenomenology of {\em jet quenching} 
--- a concept which covers the ensemble of the modifications in the properties of
a jet or of an energetic particle resulting from its interactions with a dense QCD
medium, like a quark-gluon plasma. A substantial part of the recent developments
refers to the evolution of a jet via multiple gluon emissions, as triggered by
the collisions between the constituents of the jet and those of the medium. 
This is particularly interesting since, as observed in \cite{Blaizot:2013hx},
the medium-induced gluon branchings have the potential to explain the
striking pattern of the `di-jet asymmetry', 
namely the fact that most of the energy lost by the
subleading jet is taken away by many soft hadrons propagating at
large angles w.r.t. to the jet axis. 

Within perturbative QCD at weak coupling, the medium-induced
jet evolution can be described as a classical stochastic process in which successive
branchings are quasi-independent from each other 
\cite{Blaizot:2012fh,Blaizot:2013hx,Blaizot:2013vha,Apolinario:2014csa} (see also Refs.~\cite{MehtarTani:2010ma,CasalderreySolana:2010eh,MehtarTani:2011tz,CasalderreySolana:2011rz,CasalderreySolana:2012ef} for earlier, related studies). In the most interesting physical 
regime, where the gluon radiation is triggered by multiple soft scattering, the branching rate
is given by the BDMPSZ mechanism which takes into account the coherence between successive scatterings which contribute to a single emission (a.k.a. the Landau-Pomeranchuk-Migdal effect)
 \cite{Baier:1996kr,Baier:1996sk,Zakharov:1996fv,Zakharov:1997uu,Baier:1998kq,Baier:2000sb,Wiedemann:2000za,Wiedemann:2000tf,Arnold:2001ba,Arnold:2001ms,Arnold:2002ja}. Unlike the rate
from bremsstrahlung in the vacuum, which depends only upon the splitting fraction $z$
and favors {\em soft} ($z\ll 1$) splittings, the BDMPSZ rate also depends upon the energy
$\omega$ of the parent gluon and is such that it favors {\em quasi-democratic} branchings
--- that is, $1\to 2$ branching processes for which the splitting fractions of
the daughter gluons are comparable with each other: $z\sim 1- z$.

More precisely, for a jet propagating through the medium along a distance $L$, 
the BDMPSZ mechanism introduces a characteristic medium scale, 
the {\em branching energy} $\obr(L)=\abar^2\hat q L^2$, with $\hat q$ the jet 
quenching parameter (the rate for transverse momentum broadening via elastic collisions).
The relatively soft gluons with energies $\omega\lesssim \obr(L)$ have a probability
of order one to disappear via democratic branchings and thus transmit their whole energy
--- via a mechanism akin to {\em wave turbulence} \cite{Blaizot:2013hx} ---
to a large number of very soft quanta, which are easily deviated to large angles via rescattering in
the medium. In particular, if the medium is a quark-gluon plasma in thermal equilibrium,
then the softest quanta produced by the branching process are expected to thermalize
\cite{Iancu:2015uja}.  The energy taken away by these soft quanta can be identified with
the energy lost by the jet towards the medium. In the experimental conditions at the LHC, the medium
scale $\obr(L)$ is expected to be relatively hard (of the order of a few GeV), yet significantly
softer than the original energy $E\ge 100$\,GeV of the `leading particle' (the parton
initiating the jet). In this {\em high-energy} regime at $E\gg\obr(L)$,
that will represent our main focus in that follows, the energy lost via soft quanta propagating
at large angles is controlled by wave turbulence and is proportional to $\obr(L)$ \cite{Blaizot:2013hx}.

The {\em average} picture of the medium-induced jet evolution has been studied in great
detail (at least, under suitable approximations to be later specified) 
\cite{Blaizot:2013hx,Kurkela:2014tla,Blaizot:2014ula,Fister:2014zxa,Blaizot:2014rla,Blaizot:2015jea,Iancu:2015uja,Blaizot:2015lma}. In particular,
Ref.~\cite{Blaizot:2013hx} presented exact, analytic, solutions for the gluon spectrum
$D(\omega)=\omega (\rmd N/\rmd\omega)$ and for the average energy lost by the jet at large angles.
But this stochastic process is also expected to develop {\em event-by-event fluctuations} --- 
say, in the number and the energy distribution of the branching products ---, which have been
less studied so far and which might be interesting too for the phenomenology. 

The importance of fluctuations for a parton cascade generated via $1\to 2$ gluon branchings
is demonstrated by our experience with a jet evolving in the vacuum. In that case, the branchings
are fully controlled (via the DGLAP dynamics) by the virtualities of the partons in the jet.
The statistical properties of that process --- that is, the mean gluon multiplicity
and its higher moments --- have been explicitly computed in the double-logarithmic approximation
\cite{Dokshitzer:1991wu}. One has thus discovered \cite{Bassetto:1979nt} (see also
Chapter 5 in the textbook \cite{Dokshitzer:1991wu}) the existence of large statistical fluctuations
together with a remarkable regularity known as {\em  KNO scaling} (from Koba, Nielsen, and Olesen \cite{Koba:1972ng}): in the large-virtuality limit, where the parton multiplicities are high, 
all the higher moments $\langle N^p\rangle$ with $p\ge 2$ are entirely fixed by the
average multiplicity $\langle N\rangle$. More precisely, the reduced moments 
$\langle N^p\rangle/ \langle N\rangle^p$ are pure numbers, independent of
the virtuality, explicitly known in the approximations of interest.

Returning to our actual problem, that of a jet which propagates inside 
a dense medium, one may expect the associated fluctuations to be even larger, due to the 
`democratic' nature of the medium-induced gluon branchings: after each such a splitting,
one loses trace of the parent parton, hence the ensuing cascade looks even more
`disordered' than for a jet which evolves in the vacuum. This will be confirmed and substantiated
by our subsequent findings in this paper.

On the experimental side, the extensive studies of di-jet asymmetry at the LHC revealed
that the `missing energy' is balanced by a rather large number (10 to 15) of relatively
soft hadrons (with $p_T$ between 0.5 and 2 GeV), which propagate in the hemisphere of 
the subleading jet, at large angles w.r.t. the jet axis 
\cite{Chatrchyan:2011sx,Khachatryan:2015lha,Khachatryan:2016erx}. Not surprisingly, 
this number shows large fluctuations event-by-event. 
However it seems difficult to also measure the multi-hadron
correlations (like the $p$-body moments $\langle N^p\rangle$ with $p\ge 2$), due to
the large background associated with the underlying event in a Pb+Pb collision.

On the theory side, there are only a couple of recent analyses of the importance of
statistical fluctuations for the in-medium evolution of a jet (in particular, for the di-jet asymmetry) 
\cite{Milhano:2015mng,Escobedo:2016jbm}. Ref.~\cite{Milhano:2015mng} presented
a numerical study based on the Monte-Carlo event generator JEWEL  \cite{Zapp:2012ak},
with very interesting conclusions: the di-jet asymmetry $A_J$ in central Pb+Pb collisions appears
to be controlled by fluctuations in the branching process and not by the difference
between the in-medium path lengths, $L_1$ and $L_2$, of the two jets. In fact, the typical di-jet events generated by JEWEL are such that $L_1$ and $L_2$ are rather close to each other, a situation
which according to the usual wisdom should lead to small values for $A_J$. In spite of that,
the numerical results exhibit a rather large fraction of events with large values for $A_J$, 
including for the class of events where one enforces the condition $L_1=L_2$. This clearly
demonstrates the importance of fluctuations.

In an independent, fully analytic, study which appeared soon after \cite{Escobedo:2016jbm},
we have for the first time computed the dispersion in the energy lost by the jet at large angles
and in the multiplicity of soft gluons produced via medium-induced multiple branching.
To that aim, we relied on an exact result for the gluon pair density, that we obtained
under the same assumptions as used in previous studies of the average picture
\cite{Blaizot:2013hx,Kurkela:2014tla,Blaizot:2014ula,Fister:2014zxa,Blaizot:2014rla,Blaizot:2015jea,Blaizot:2015lma}. Our results demonstrate that the fluctuations are huge: for both quantities alluded
to above, the dispersion is parametrically as large as the respective mean value.
In particular, if $\mcal{E}$ denotes the energy lost by the jet event by event, then in the high-energy
regime at $E\gg\obr$, we found that 
$\sigma_{\mcal{E}}\sim \langle\mcal{E}\rangle \sim \obr$, where 
$\sigma_{\mcal{E}}^2\equiv \langle\mcal{E}^2\rangle-\langle\mcal{E}\rangle^2$.
(We recall that $\obr(L)=\abar^2\hat q L^2$ is the characteristic medium scale for
multiple branching and $E$ is the initial energy of the jet.) This in turn implies that
the fluctuations in the medium-induced branching process can contribute to the di-jet
asymmetry\footnote{With the present notations, the di-jet asymmetry that is actually
measured at the LHC can be written as $A_J=({E_>-E_<})/({E_1+E_2})
= |\mcal{E}_1-\mcal{E}_2|/({2E-\mcal{E}_1-\mcal{E}_2})$, where $E_i=E-\mcal{E}_i$
are the final energies of the 2 jets, $E$ is their common initial energy, and
$E_>$ ($E_<$) is the largest (smallest) among $E_1$ and $E_2$. Hence $A_J$ is
by definition semi-positive and should be compared to 
$\langle (\mcal{E}_2-\mcal{E}_1)^2\rangle $, and {\em not} to the average
difference $\langle \mcal{E}_2-\mcal{E}_1\rangle$, which can have either sign.}
at the same level as the difference in path-lengths between the two jets:
\beq\label{sigmaAJ}
\langle (\mcal{E}_2-\mcal{E}_1)^2\rangle = \langle \mcal{E}_2-\mcal{E}_1\rangle^2+
\sigma_{\mcal{E}_1}^2+\sigma_{\mcal{E}_2}^2\,
\propto\big(L_1^2-L_2^2\big)^2 + \big(L_1^4+L_2^4\big)\,.
\eeq
The first term in the r.h.s., proportional to the difference $L_1^2-L_2^2$,
is the average contribution $\langle \mcal{E}_2-\mcal{E}_1\rangle$ and
vanishes when $L_1=L_2$, as expected. But the second term, originating 
from fluctuations, is always nonzero and it dominates over the average piece
whenever $L_1$ and $L_2$ are close to each other.

Clearly, the findings in Refs.~\cite{Milhano:2015mng,Escobedo:2016jbm} are
consistent with each other, the reinforce each other and may together
lead to a shift of paradigm concerning the physical origin of the di-jet asymmetry.

In this paper, we shall complete the study, started in Ref.~\cite{Escobedo:2016jbm},
of multi-particle correlations in the gluon distribution produced via multiple medium-induced branchings. Our main new result is a set of exact, analytic, expressions for the 
$p$-body densities $\mcal{N}^{(p)}(x_1,\cdots,x_p)$ which describe the gluon distribution
in the space of energy. Here, $x_i\equiv \omega_i/E$ is the energy fraction of one of the
$p$ measured gluons w.r.t. the leading particle. The respective expressions for  $p=1$
[the gluon density $\mcal{N}(x)$, which is related to the spectrum $D(x)$ via
$D(x)=x\mcal{N}(x)$] and $p=2$ [the gluon pair density $\mcal{N}^{(2)}(x_1,x_2)$] were
already known, as previously mentioned, but those for the higher-point correlations
with $p\ge 3$ are new. The 3-gluon density $\mcal{N}^{(3)}$ is shown
in \eqn{N3exact} and the generic $p$-body function $\mcal{N}^{(p)}$ 
in Eqs.~\eqref{Npexact}--\eqref{hp}. By inspection of these explicit results and of their
derivation, one uncovers generic features and structural properties which shed more light
on the physical picture for the in-medium jet evolution. For instance, all these correlations
exhibit a `geometric scaling' property: they depend upon the physical parameters 
$\alpha_s$, $\hat q$, $L$, and $E$ only via the dimensionless ratio $\obr(L)/E$.

The emerging physical picture for a typical event can be summarized as follows\footnote{We 
consider here the high-energy regime at $E\gg\obr(L)$, which is the most interesting one
for the phenomenology at the LHC. For the corresponding picture at lower energies, $E\lesssim
\obr(L)$, see the discussion in Sect.~\ref{sec:sols}.}:
the jet is structured as an ensemble of `mini-jets', i.e. gluon cascades generated via
successive democratic branchings by  `primary partons' (gluons directly emitted
by the leading particle) with relatively low energies $\omega\le \obr(L)$. Harder primary 
emissions, with $\obr(L) \ll \omega < E$, are possible as well, but they occur with a low 
probability (i.e. only in rare events) and do not give rise to mini-jets 
(since hard gluons cannot suffer democratic branchings). All the partons from a mini-jet are
strongly correlated with each other, as they have a common ancestor to which
they are linked via a succession of democratic branchings. On the other hand, 
different mini-jets are uncorrelated with each other, since the successive emissions
of soft primary gluons are quasi-independent (indeed, the constraint of
energy conservation plays only a minor role for the soft emissions).

By integrating the multi-gluon densities $\mcal{N}^{(p)}(x_1,\cdots,x_p)$ 
over their energy arguments $x_i$,
above some suitable infrared cutoff $x_0\equiv \omega_0/E$, we shall deduce the
{\em gluon multiplicities} --- the average number $\langle N\rangle(\omega_0)$ of
gluons with energies $\omega\ge\omega_0$ together with its higher moments\footnote{More
precisely, we shall compute the {\em factorial} moments 
$\langle N(N-1) \dots (N-p+1)\rangle$, but this distinction is unimportant in the
high-multiplicity regime of interest, where $N\gg p$.}  $\langle N^p\rangle(\omega_0)$.
The lower cutoff $\omega_0$ plays the role of an energy resolution scale. Without such a cutoff,
the gluon multiplicity would be divergent, due to the
rapid rise in the emission probability as $\omega\to 0$. On more physical grounds, one should
observe that the `ideal' (or `turbulent') branching picture under consideration holds only
for sufficiently high energies $\omega\gg T$, with $T$ the average $p_T$ of the medium constituents (say, the temperature if the medium is a quark-gluon plasma). Hence, a physically meaningful
cutoff satisfies $\omega_0\sim T \ll \obr(L)\ll E$. With this choice, the multiplicities are
parametrically large and dominated by the softest measured gluons,
those with energies $\omega\sim\omega_0$. This last feature allows for relatively simple
analytic approximations (see Sect.~\ref{sec:number} for details).

Specifically, we shall find that 
$\langle N\rangle \sim (\obr/\omega_0)^{1/2} \gg 1$ and $\langle N^p\rangle
\sim \langle N\rangle^p$. This in particular implies {\em KNO scaling}: the reduced moments 
$\kappa^{(p)}\equiv \langle N^p\rangle/ \langle N\rangle^p$ are pure numbers, independent of
all the physical parameters of the problem, i.e. $\alpha_s$, $\hat q$, $L$, and $E$ (see 
\eqn{kappapsmall}). This feature is also shown by a jet evolving in the vacuum
\cite{Dokshitzer:1991wu}, but the respective distributions are significantly different: 
the statistical fluctuations are considerably larger for the medium-induced evolution.
A precise way to characterize this difference is by comparing the corresponding probability
distributions $\mathcal{P}(N)$, which in the KNO regime are fully specified by the
set of numbers $\kappa^{(p)}$ with $p\ge 1$. For the medium-induced evolution,
we shall recognize $\mathcal{P}(N)$ as a specific {\em negative
binomial distribution} (NBD) \cite{Feller}, 
that with parameter $r=2$ (see Sect.~\ref{sec:NBD} for details). This distribution is
indeed broader (in the sense of developing larger fluctuations) than the one generated by
a jet in the vacuum; the latter is known only numerically and can be viewed as an
interpolation between the two NBD's with $r=2$ and $r=3$, respectively \cite{Dokshitzer:1991wu}.

The emergence of a NBD in relation with the medium-induced jet evolution is
perhaps surprising, at the same level as the existence of exact, analytic, solutions
for all the multi-gluon correlations. Notice indeed that we have a better analytic control
for the evolution of the jet in the medium than in the vacuum, albeit the latter is
{\em a priori} supposed to be a simpler problem.
Without having a fully convincing explanation in that sense,
we believe that both features could be related to the physics of wave turbulence,
more precisely, to the existence of fixed-point solutions to the evolution equations for
the multi-gluon densities $\mcal{N}^{(p)}(x_1,\cdots,x_p)$. On one hand, these fixed
points greatly facilitate the search for analytic solutions. On the other hand, they determine
the multi-gluon spectra at low energies $x\ll 1$ and hence, in particular, the multiplicities
of soft gluons.

\comment{So far, we have a rather good understanding, analytic and fully explicit, of the stochastic
aspects of the parton cascades produced either by a jet propagating through
the vacuum, or by a jet which interacts with a dense medium, but
such that the jet constituents are nearly on-shell  (meaning that all radiative
processes are triggered by interactions in the medium). It is of utmost importance, 
notably in view of applications to phenomenology, to complete this picture for the in-medium jet
evolution by adding the parton virtualities and the associated vacuum-like radiation.
This would most likely require to abandon a purely analytic approach, but rely on
numerical methods like the Monte-Carlo event generators 
JEWEL  \cite{Zapp:2012ak} and MARTINI  \cite{Schenke:2009gb}.
}

This paper is organized as follows. In Sect.~\ref{sec:Master} we briefly review the 
theoretical description of the medium-induced gluon branching as a Markovian stochastic
process. In this context, we shall present for the first time the evolution equation obeyed
by the $p$-body density $\mcal{N}^{(p)}(x_1,\cdots,x_p)$ with generic $p\ge 1$. 
More details on the construction of this equation are presented in Appendix~\ref{app}.
In Sect.~\ref{sec:sols} we present and discuss our exact results for the multi-gluon
correlations $\mcal{N}^{(p)}(x_1,\cdots,x_p)$. We first recall the known results for $p=1$
and $p=2$ (but our physical discussion of the pair density $\mcal{N}^{(2)}$
in Sect.~\ref{sec:pair} is largely new). Then we present the new results for the 3-point
function $\mcal{N}^{(3)}$ (in Sect.~\ref{sec:3point}) and the general $p$-point function
(in Sect.~\ref{sec:3point}).  The recursive construction of $\mcal{N}^{(p)}$ is described in
more detail in Appendix \ref{sec:appNN}. Some limiting forms of the general result for
$\mcal{N}^{(p)}$, as exhibited in Sect.~\ref{sec:3point}, will be explicitly derived in
Appendices \ref{sec:smalll} and \ref{sec:largel}. Sect.~\ref{sec:mult} is devoted to a
study of the gluon multiplicities. These are defined and computed (modulo some approximations)
in Sect.~\ref{sec:number}. Then in Sect.~\ref{sec:KNO} we discuss the KNO scaling
and in Sect.~\ref{sec:NBD} the associated, negative-binomial, distribution. Finally,
Sect.~\ref{sec:conc} presents our conclusions together with a brief
discussion of the limitations of the present formalism and some open problems.

\section{Master equations}
\label{sec:Master}

We consider the parton cascade --- to be subsequently referred as `the jet', for brevity ---
which is generated via multiple gluon branchings by an incoming parton --- the `leading particle' (LP) --- 
with initial energy $E$ which crosses the medium along a distance (or time) $L$ (the `medium size').
We assume the LP to be on-shell at the time when it enters the medium, so that all the
subsequent branchings are induced by its interactions, and the interactions of its descendants, 
with the constituents of the medium. We furthermore assume the medium to be a weakly coupled
quark-gluon plasma in thermal equilibrium with temperature $T\ll E$. The most important gluon emissions
for what follows are those with intermediate energies, within the range $T\ll \omega\ll \hat q L^2$,
for which the formation times are much smaller than the medium size $L$, but much larger than
the average mean free path between two successive collisions in the plasma. The rate for
such medium-induced emissions can be computed in the multiple soft scattering approximation,
with a result known as the BDMPSZ spectrum for a single gluon emission
 \cite{Baier:1996kr,Baier:1996sk,Zakharov:1996fv,Zakharov:1997uu,Baier:1998kq}.
Moreover, as shown in \cite{Blaizot:2012fh,Blaizot:2013vha,Apolinario:2014csa}, 
successive emissions can be treated as independent from each other,
because the typical duration between two emissions (the `branching time' to be introduced in
\eqn{tbr} below) is parametrically larger than the formation time for individual emissions and,
moreover, the coherence between the daughter gluons is efficiently washed out by the 
scattering in the medium.
As a result, the jet evolution via medium-induced gluon branching can be described as a 
{\em Markovian stochastic process} \cite{Blaizot:2013vha,Escobedo:2016jbm}.

Our goal throughout this paper is to study the energy distribution generated by this stochastic process, 
including fluctuations and correlations. To characterize this distribution, 
we shall compute the {\em factorial moment densities} $\mcal{N}^{(p)}(x_1,x_2,\dots,x_p|\tau)$
for any $p\ge 1$ and for generic values for the energy fractions $x_i\equiv \omega_i/E$
and for the `reduced time' $\tau$. The factorial moment density  $\mcal{N}^{(p)}$ is roughly speaking
the $p$-body density in the space of energy; this will be more precisely defined in \eqn{Np}
below. The reduced time is defined as 
$\tau\equiv t/\tbr(E)$, where $t\le L$ is the actual time (or distance) 
travelled by the leading particle across the medium and the reference scale $\tbr(E)$ is 
the  `democratic branching time' for the LP --- that is, the `lifetime' of the LP until it
disappears via a quasi-democratic branching. By `quasi-democratic' we mean a $1\to 2$
gluon branching where the daughter gluons carry comparable fractions of the energy of
their parent parton. For a parent gluon with energy
$\omega$, one has (see e.g. the discussion in \cite{Escobedo:2016jbm})
\beq\label{tbr}
\tbr(\omega)\,= \,\frac{1}{\abar}\sqrt{\frac{\omega} {\hat q}}\,,\eeq
with $\abar\equiv \alpha_s N_c/\pi$ ($\alpha_s$ is the QCD coupling, assumed 
to be fixed, and $N_c$ is the number of colors) and
$\hat q$ the `jet quenching parameter' (the transport coefficient for transverse 
momentum diffusion). 

We have anticipated here that, to the approximations of interest, 
the gluon distribution produced by the medium-induced jet evolution
shows an interesting, {\em geometric scaling}, property: for given values of
the energy fractions $x_i$, the factorial moment densities
$\mcal{N}^{(p)}(x_1,x_2,\dots,x_p|\tau)$ depend upon the physical parameters
of the problem --- the travelled distance $L$, the transport coefficient $\hat q$, and
the original energy $E$ of the LP --- via a single, dimensionless, variable:
 the reduced time $\tau=\abar L\sqrt{{\hat q}/{E}}$.  Accordingly, the gluon correlations are
not modified when simultaneously changing, say, the medium size $L$ and the energy $E$,
but in such a way to keep constant the ratio $L^2/E$.

The jet evolution via medium-induced multiple branching is a stochastic process 
whose dynamics is most economically expressed
in terms of  the {\em generating functional}
 \beq\label{Zdef}
Z_\tau[u(x)]\,\equiv\,\sum_{n=1}^{\infty}\int \prod_{i=1}^n \rmd x_i\,\mcal{P}_n(\{x\}|\tau)\,
 u(x_1)u(x_2)\dots u(x_n)\,,\eeq
with $ \mcal{P}_n(x_1,x_2,\cdots,x_n| \tau)$ the probability density 
for having a state with $n$ gluons with energy fractions $x_i$ ($i=1,\dots,n$), at 
(reduced) time $\tau$ and $u(x)$ an arbitrary `source' function with support at $0\le x\le 1$. 
At $\tau=0$, we have just the LP, hence $ \mcal{P}_n(\tau=0)=\delta_{n1}\delta(x_1-1)$.
Probability conservation requires $Z_\tau[u=1]=1$ for any $\tau \ge 0$.
The expectation value of an arbitrary observable is computed as
\beq\label{aveO}
\langle \mcal{O}(\tau)\rangle\equiv
\sum_{n=1}^\infty\int \prod_{i=1}^n \rmd x_i\,  \mcal{P}_n(x_1,x_2,\cdots,x_n| \tau)\,
\mcal{O}_n\,,\eeq
where $\mcal{O}_n\equiv \mcal{O}(x_1,x_2,\cdots,x_n)$ denotes the value of
$\mcal{O}$ in a particular event with $n$ gluons. 

Strictly speaking, this probabilistic description requires an infrared 
cutoff (say, a lower limit on $x$), playing the role of an energy resolution scale, below which
gluons cannot be resolved anymore. Indeed, the branching dynamics produces an
infinite number of arbitrarily soft gluons and the `state with exactly $n$ gluons' is not well defined
without such a cutoff.  Any explicit construction of such a state, say via Monte-Carlo 
simulations, must involve an infrared cutoff on $x$, to be viewed as a part of the `state' definition.
On the other hand, the correlation functions of interest, like the $p$-body densities
$\mcal{N}^{(p)}(x_1,x_2,\dots,x_p|\tau)$, are insensitive to the unobserved, soft, gluons, 
hence they are independent of this cutoff. So long as one is solely interested in such 
semi-inclusive quantities, one can formally proceed without introducing any infrared cutoff.

Given the generating functional \eqref{Zdef}, the $p$-body densities of interest are
obtained via functional differentiation w.r.t. the source field $u(x)$~:
\beq\label{Np}
\mcal{N}^{(p)}(x_1,x_2,\dots,x_p|\tau)\,=\,
\frac{\delta^p Z_\tau[u]}{\delta u(x_1)\delta u(x_2)\dots \delta u(x_p)}\bigg|_{u=1}\,.
 \eeq
In particular, for $p=1$ one finds the gluon density in $x$-space, or gluon spectrum, and for $p=2$,
the density of pairs of gluons (with each pair being counted twice):
\beq\label{N12}
\mcal{N}(x,\tau)\equiv \mcal{N}^{(1)}(x|\tau)
=\left\langle \sum_i^n\delta(x_i-x)\right\rangle\,,\qquad 
\mcal{N}^{(2)}(x,x'|\tau)= \left\langle 
\sum_{i\ne j}^n\delta(x_i-x) \delta(x_j-x')\right\rangle\,.\eeq
The function $\mcal{N}^{(p)}(x_1,x_2,\dots,x_p|\tau)$ with $p\ge 2$ is totally symmetric
under the permutations of the variables $x_i$.

The  {\em factorial cumulant densities}, obtained from the logarithm of the
generating functional,
\beq\label{Cp}
\mcal{C}^{(p)}(x_1,x_2,\dots,x_p|\tau)\,=\,
\frac{\delta^p \ln Z_\tau[u]}{\delta u(x_1)\delta u(x_2)\dots \delta u(x_p)}\bigg|_{u=1}\,.
 \eeq
will be useful too, as they measure genuine correlations in the gluon distribution. In
particular,
\begin{align}\label{C23}
\mcal{C}^{(2)}(x_1,x_2|\tau)=&\,\mcal{N}^{(2)}(x_1,x_2|\tau)-
\mcal{N}(x_1,\tau)\mcal{N}(x_2,\tau)\,.\nn
\mcal{C}^{(3)}(x_1,x_2,x_3|\tau)=&\,\mcal{N}^{(3)}(x_1,x_2,x_3|\tau)-
\mcal{N}^{(2)}(x_1,x_2|\tau)\mcal{N}(x_3,\tau)-\mcal{N}^{(2)}(x_1,x_3|\tau)\mcal{N}(x_2,\tau)
\nn &-\mcal{N}^{(2)}(x_2,x_3|\tau)\mcal{N}(x_1,\tau)
+2\mcal{N}(x_1,\tau)\mcal{N}(x_2,\tau)\mcal{N}(x_3,\tau).
\end{align}
 
The time evolution of all the $p$-body correlations is succinctly described 
by a single, functional, evolution
equation for $Z_\tau[u]$, which reads  \cite{Escobedo:2016jbm}
  \beq\label{eqZ}
  \frac{\del Z_\tau[u]}{\del\tau}\,=\,
  \int\rmd z \int\rmd x\,\mcal{K}(z,x)\big[u(zx) u((1-z)x)- u(x)\big]\,
   \frac{\delta Z_\tau[u]}{\delta u(x)}\,,\eeq 
 where the kernel (the `reduced' version of the BDMPSZ spectrum)
  \beq\label{Kzx}
  \mcal{K}(z,x)= \,\frac{\mcal{K}(z)}{2\sqrt{x}}\qquad\mbox{with}\qquad
 {\cal K}(z)\equiv\frac{[1-z(1-z)]^{\frac{5}{2}}}{[z(1-z)]^{\frac{3}{2}}}={\cal K}(1-z)\,,
  \eeq
is the differential probability per unit (reduced) time and per unit $z$ for the splitting
of a parent gluon with energy fraction $x$ into a pair of daughter gluons with energy
fractions $zx$ and $(1-z)x$, with $0 < z < 1$ (the splitting fraction). The functional derivative 
${\delta Z_\tau[u]}/{\delta u(x)}$ in the r.h.s. of \eqn{eqZ}
plays the role of an `annihilation operator' (it reduces
by one the number of factors of $u$). The term quadratic in $u$ within the square brackets
describes the `gain'
effect associated with the branching process $x\to (zx, (1-z)x)$, whereas the negative
term linear in $u$ is the corresponding `loss' effect.

By taking $p$ functional derivatives in \eqn{eqZ} and evaluating the result 
at $u(x)=1$, it is straightforward to obtain the evolution equation obeyed by 
the $p$-th factorial moment. The respective equations for $p=1$ and $p=2$ have 
been presented in previous publications 
\cite{Blaizot:2013hx,Escobedo:2016jbm}, but will be repeated here, for more clarity
 (and also because our present conventions are slightly different). They read
  \beq\label{eqN}
\frac{\partial}{\partial\tau} \,\mcal{N}(x,\tau)=\frac{1}{\sqrt{x}}\int \rmd z \,
{\cal K}(z)\left[\frac{1}{\sqrt{z}}\, \mcal{N}\left(\frac{x}{z},\tau\right)-z\,\mcal{N}(x,\tau)\right],
\eeq
and respectively 
\begin{align}\label{eqN2}
   \frac{\del }{\del\tau}\,\mcal{N}^{(2)}(x_1,x_2|\tau)
  &=\frac{1}{\sqrt{x_1}}\int \rmd z \,
  {\cal K}(z)\left[\frac{1}{\sqrt{z}} \,\mcal{N}^{(2)}\Big(\frac{x_1}{z}, x_2\big |\tau\Big)- z \,
  \mcal{N}^{(2)}\big({x_1},x_2|\tau\big)\right]
  \, + \,\big( x_1 \,\leftrightarrow\, x_2\big)\nonumber\\*[0.2cm]
  &\qquad +\,\frac{1}{(x_1+x_2)^{3/2}}\,
   {\cal K}\Big(\frac{x_1}{x_1+x_2}\Big)\,\mcal{N}(x_1+x_2,\tau)\,.
  \end{align}

\eqn{eqN} is homogeneous and must be solved with the initial condition 
$\mcal{N}(x,\tau=0)=\delta(x-1)$. The first term in its r.h.s. describes the gain in the number 
of gluons at  $x$ due to emissions from gluons with $x'=x/z > x$, 
whereas the second term describes the loss via the decay into softer gluons.

\eqn{eqN2} is inhomogeneous; its r.h.s. involves the source term
\beq\label{S2}
S^{(2)}(x_1,x_2|\tau)\equiv \,\frac{1}{(x_1+x_2)^{3/2}}\,
   {\cal K}\Big(\frac{x_1}{x_1+x_2}\Big)\,\mcal{N}(x_1+x_2,\tau)\,,\eeq
which describes the simultaneous creation of a pair of gluons with energy
fractions $x_1$ and $x_2$ via the branching of a parent gluon with energy fraction $x_1+x_2$
(with $x_1+x_2 \le 1$ of course). After this splitting, the two daughter gluons evolve independently
from each other and create their own gluon distributions (this evolution is described by
the first line of \eqn{eqN2}). Accordingly, the solution $\mcal{N}(x,\tau)$  to \eqn{eqN} 
acts as a Green's function for  \eqn{eqN2}: the solution to the latter with the initial condition
$\mcal{N}^{(2)}(x_1,x_2|\tau=0)=0$ can be written as
 \begin{align}\label{N2sol}
   \mcal{N}^{(2)}(x_1,x_2|\tau)
   =\int_0^\tau\rmd \tau'\int^1_{x_1}\frac{\rmd \xi_1}{\xi_1}\int^{1-\xi_1}_{x_2}\frac{\rmd \xi_2}{\xi_2}\,
      \mcal{N}\bigg(\frac{x_1}{\xi_1},\frac{\tau-\tau'}{\sqrt{\xi_1}}\bigg) \,
      \mcal{N}\bigg(\frac{x_2}{\xi_2},\frac{\tau-\tau'}{\sqrt{\xi_2}}\bigg)S^{(2)}(\xi_1,\xi_2|\tau')\,, \end{align}
with a transparent physical interpretation: at some intermediate time $\tau'$, a gluon with energy 
fraction $\xi_1+\xi_2$ splits into two gluons with energy fractions $\xi_1$ and respectively $\xi_2$,
whose subsequent evolutions generate two mini-jets which include the final gluons, 
$x_1$ and respectively $x_2$. 
Note that the parent gluon with energy $\xi_1+\xi_2$ is the {\em last common ancestor} (LCA) of the
two measured gluons $x_1$ and $x_2$.

The equation obeyed by $\mcal{N}^{(p)}$ for generic $p\ge 1$ will be derived
in Appendix~\ref{app} and reads
\begin{eqnarray}\label{eqNp}
&&\frac{\partial }{\partial\tau}\,\mcal{N}^{(p)}(x_1,\cdots,x_p|\tau)=\nonumber\\
&&\qquad\sum_{i=1}^p
\frac{1}{\sqrt{x_i}}\int \rmd z \,
  {\cal K}(z)\left[\frac{1}{\sqrt{z}} \,\mcal{N}^{(p)}\Big(x_1,\cdots,\frac{x_i}{z},\cdots,x_p\big |\tau\Big)- z \,
  \mcal{N}^{(p)}\big(x_1,\cdots,x_p|\tau\big)\right]
\nonumber\\
&&\qquad+\sum_{i=2}^p\sum_{j=1}^{i-1}\frac{1}{(x_i+x_j)^{3/2}}\mcal{K}\left(\frac{x_i}{x_i+x_j}\right)
\,\mcal{N}^{(p-1)}(x_1,\cdots,x_i+x_j,\cdots,x_p|\tau)\,.
\end{eqnarray}
The similarity with \eqn{eqN2} is quite manifest: the distribution $\mcal{N}^{(p-1)}$ for $p-1$
particles acts as a source for the $p$-body density $\mcal{N}^{(p)}$. Specifically, the source term
in the above equation describes the {\em simultaneous} creation of the pair of particles $x_i$ and $x_j$
(taken among the $p$ particles measured by $\mcal{N}^{(p)}$) via the splitting of one gluon
with energy fraction $x_i+x_j$ that was included in $\mcal{N}^{(p-1)}$.  \eqn{eqNp} is
formally solved by (with the initial condition $\mcal{N}^{(p)}(\tau=0)$)
\begin{equation}
\mcal{N}^{(p)}(x_1,\cdots,x_p|\tau)=\int_0^\tau\,\rmd\tau'\prod_{i=1}^p\int_{x_i}^1\frac{\,\rmd\xi_i}{\xi_i}\,
\mcal{N}\left(\frac{x_i}{\xi_i},\frac{\tau-\tau'}{\sqrt{\xi_i}}\right)\Theta\bigg(1-\sum_{j=1}^p\xi_j\bigg)
S^{(p)}(\xi_1,\cdots,\xi_p|\tau')\,,
\label{Npsol}
\end{equation}
where $S^{(p)}(\xi_1,\cdots,\xi_p|\tau')$ denotes the source term in the r.h.s. of  \eqn{eqNp}.
\eqn{Npsol} is truly a recursion formula, which expresses the $p$-body density $\mcal{N}^{(p)}$
in terms of the $(p-1)$-th one.

\section{Analytic solutions for the multi-gluon correlations}
\label{sec:sols}

From now on, we shall focus on a slightly simplified version of the master equations
introduced in the previous section, which has the virtue to allow for explicit, analytic, solutions,
while at the same time keeping all salient features of the dynamics. This version is obtained by 
replacing the original kernel ${\cal K}(z)$ (i.e. the branching rate) 
with ${\cal K}_0(z)\equiv 1/{[z(1-z)]^{3/2}}$. The simplified kernel preserves the pole
structure of the exact kernel at $z=0$ and $z=1$, hence it generates a very similar evolution. 
This is confirmed by numerical solutions to \eqn{eqN} using both forms of the kernel 
\cite{Fister:2014zxa,Blaizot:2015jea}.

For this simplified kernel, one was able to obtain exact, analytic solutions 
for the gluon spectrum $\mcal{N}(x,\tau)$ \cite{Blaizot:2013hx} and the pair density
$\mcal{N}^{(2)}(x_1,x_2|\tau)$ \cite{Escobedo:2016jbm}. In what follows, we shall first
briefly review these known solutions and thus introduce a physical picture for the
medium-induced jet evolution which will later be refined by our new results
for the higher $p$-point functions.

\subsection{The gluon spectrum}
\label{sec:spectrum}

The gluon spectrum $\mcal{N}(x,\tau)$ corresponding to the 
simplified kernel ${\cal K}_0(z)$ reads  \cite{Blaizot:2013hx}
\beq\label{Nexact}
  \mcal{N}(x,\tau)\,=\,\frac{\tau}{[x(1-x)]^{3/2}}\,\rme^{-\frac{\pi\tau^2}{1-x}},\eeq  
For relatively small times $\pi\tau^2\ll 1$, i.e. $t\ll \tbr(E)$, this spectrum
exhibits a pronounced peak near $x=1$ which describes the leading particle together with a power
tail $\mcal{N}(x,\tau)\simeq \tau/x^{3/2}$ at $x\ll 1$, which describes soft radiation.
The shift $1-x_p\sim \pi\tau^2$ in the position of the peak measures the typical energy lost by the LP
via radiation, where the width of this peak $\delta x_p\sim\pi \tau^2$ describes the `broadening', 
i.e. the uncertainty in the energy of the LP due to fluctuations in the radiation process. In physical
units, with the notations $\Delta E_p\equiv E(1-x_p)$ and $\delta E_p\equiv E \delta x_p$, one
finds, parametrically,
\beq\label{omegabr}
\Delta E_p\,\sim\, \delta E_p\,\sim\,\omega_{\rm br}(t)\equiv \tau^2 E=\abar^2\hat q t^2\,.\eeq
This `branching energy' $\omega_{\rm br}(t)$, which is independent of $E$ and much smaller 
than it (remember that we are in the regime where  $\pi\tau^2\ll 1$), is the characteristic energy
scale for the onset of {\em multiple branching}~: there is a probability of order one to emit a
gluon  with energy $\sim \omega_{\rm br}(t)$ during a time interval $t$.
In particular, a {\em quasi-democratic} branching occurs with probability of $\order{1}$
during $t$ provided the energy $\omega_{\rm br}(t)$ of one (any) of the daughter
gluons is comparable to the energy $\omega$ of their parent; this condition
$\omega\sim \omega_{\rm br}(t)$ implies $t\sim\tbr(\omega)$, with $\tbr(\omega)$ 
introduced in \eqn{tbr}.

The estimates in \eqn{omegabr} can be physically understood as follows \cite{Escobedo:2016jbm}:
during the relatively small time $t\ll \tbr(E)$, the LP cannot undergo a democratic branching, rather
it radiates a large number of very soft gluons with energies $\omega \ll \omega_{\rm br}(t)$ 
together with a number of $\order{1}$
of harder gluons with $\omega \sim \omega_{\rm br}(t)$. The latter control
the energy lost by the LP in a typical event, hence $\Delta E_p(t)\sim \omega_{\rm br}(t)$.
Besides, the {\em fluctuations} in the number of hard gluons are of $\order{1}$ as well
(since successive hard emissions are quasi-independent),
hence also the broadening $\delta E_p(t)$ must be of order $\omega_{\rm br}(t)$.

After being emitted by the LP, the primary gluons with energies $\omega \lesssim \omega_{\rm br}(t)$ 
are bound to undergo democratic branchings, because the corresponding branching times obey
$\tbr(\omega)  \lesssim t < L$. Via successive  democratic branchings, they generate 
parton cascades (`mini-jets') and thus gradually transfer their energy to softer and softer
quanta, and eventually to the medium. This mechanism for energy transfer is extremely efficient,
since characterized by {\em wave turbulence}   \cite{Blaizot:2013hx}:  the rate for energy flow
from one parton generation to the next one is independent of the generation (i.e. of $x$).
This is visible in the special form of the gluon spectrum \eqref{Nexact} at small $x$, namely
$\mcal{N}(x)\propto x^{-3/2}$ : this power law represents a {\em turbulent fixed point} 
for the branching process. That is, the gain and loss terms in the r.h.s. of \eqn{eqN} mutually
cancel for this particular spectrum, meaning that there is no net accumulation of energy at any
intermediate values of $x$: after a time of order $\tbr(\omega)$, the whole energy $\omega$
that was initially carried by a primary gluon ends up into arbitrarily soft quanta ($x\to 0$). 

More precisely, this `turbulent' branching picture applies so long as the gluon energies
remain larger than the temperature of the medium: the very soft gluons with energies  
of order $T$ can efficiently exchange energy and momentum with the medium constituents, 
via elastic collisions, and thus thermalize \cite{Iancu:2015uja}. As a result, the energy $\omega$
of the primary gluon is ultimately transmitted {\em to the medium}, in the form of 
many soft quanta which emerge at large angles.

This discussion shows that the energy transfer from the LP to the medium via democratic
cascades is {\em quasi-deterministic}. Accordingly, the statistics of the energy lost {\em by the LP}, as
encoded in \eqn{omegabr}, also applies to the energy lost {\em by the jet as a whole}, via
soft radiation at large angles: both the average energy loss and its dispersion are of 
order $\omega_{\rm br}(t)$, since controlled by the hardest `primary' emissions
which can occur in a typical event \cite{Escobedo:2016jbm}. 
The above discussion applies so long as $t\ll \tbr(E)$, which is the relevant
situation for the phenomenology of di-jet asymmetry at the LHC
 \cite{Aad:2010bu,Chatrchyan:2011sx}: the initial energy $E$ of the LP is
so high ($E\ge 100$\,GeV) that the corresponding branching time $\tbr(E)$ is much 
larger than the available medium size $L\lesssim 6$\,fm.

For completeness, let us also consider the situation at larger times $t\gtrsim \tbr(E)$, 
or $\tau \gtrsim 1$. This applies to jets with a smaller overall energy $E$ and also
to the `mini-jets' generated by primary gluons, as previously discussed.
When $\pi\tau^2\sim 1$, the exponent in \eqn{Nexact} becomes of $\order{1}$ for any value
of $x$, hence the peak disappears from the spectrum: as expected, the LP undergoes
a democratic branching. When further increasing $\tau$, the support of the spectrum shrinks 
towards small values $x\lesssim 1/\pi\tau^2\ll 1$, while its strength is rapidly suppressed, as 
$\rme^{-{\pi\tau^2}}$. This confirms that all the gluons with $x > 0$ (more precisely, 
$x> T/E$) disappear via democratic branchings.

\comment{
The gluon number density is strongly divergent as $x\to 0$, but the energy density
$x\mcal{N}(x,\tau)$ is integrable and the total energy fraction contained in the spectrum 
(i.e. in the bins with $x>0$) can be easily computed as $\int_0^1 \rmd x \,x\,\mcal{N}(x,\tau)
=\rme^{-{\pi\tau^2}}$  \cite{Blaizot:2013hx}. For $\tau\gtrsim 1$, i.e. $t\gtrsim \tbr(E)$,
this is rapidly decreasing with time: the energy disappears from the spectrum.
This is another consequence of wave turbulence: 
in the absence of any infrared cutoff, the energy accumulates in a 
condensate at $x=0$. 

This overall process
--- energy degradation into soft quanta via democratic branchings followed by the thermalization
of the soft branching products --- represents a very efficient mechanism for
transferring energy from the jet to the medium. Since carried by soft quanta,
the energy that is lost in this way emerges at large angles w.r.t. the jet axis, in agreement
with the phenomenology of di-jet asymmetry at the LHC \cite{Aad:2010bu,Chatrchyan:2011sx}.
}

\subsection{The gluon pair density}
\label{sec:pair}

The gluon pair density has been computed in \cite{Escobedo:2016jbm}, with the following
result (a streamlined derivation of this result will be given below, in Appendix \ref{sec:appN})
\begin{align}\label{N2exact}
    \mcal{N}^{(2)}(x_1,x_2|\tau)\,=\,\frac{1}{2\pi}\frac{1}{\sqrt{(x_1x_2)^{3}(1-x_1-x_2)}}
    \left[\rme^{-\frac{\pi\tau^2}
    {1-x_1-x_2}}
   -\rme^{-\frac{4\pi\tau^2}{1-x_1-x_2}}\right].
   \end{align}
To better appreciate the physical interpretation of this result, it is useful to have a look
at its derivation. The difference of two exponentials in the r.h.s. has been generated
via the following integral,
\beq\label{int2p}
\int_0^\tau\,\rmd\tau'(2\tau-\tau')\,\rme^{-\frac{\pi(2\tau-\tau')^2}{1-x_1-x_2}}\,=\,\frac{1-x_1-x_2}{2\pi}
\left[\rme^{-\frac{\pi\tau^2}
    {1-x_1-x_2}}
   -\rme^{-\frac{4\pi\tau^2}{1-x_1-x_2}}\right],
\eeq
where the integration variable $\tau'$ has the same meaning as in \eqn{N2sol}:
this is the splitting time for the last common ancestor (LCA). The integrand 
in \eqn{int2p} happens to be a total derivative w.r.t. $\tau'$, hence the result of the
integration comes from the two endpoints. The first term in the r.h.s. is generated
by the upper limit $\tau'=\tau$ and describes a
process in the which the splitting of the LCA occurs very late, near the time of
measurement; the respective exponential is recognized as the
probability for the LCA (with energy fraction $x_1+x_2$) to survive
over a time $\tau$ (compare to \eqn{Nexact}). The second exponential similarly 
refers to an early splitting ($\tau'=0$) and can be interpreted as the probability that both
daughter gluons survive until they are finally measured at time $\tau$.
For relatively large measurement times, $\pi\tau^2\gtrsim 1$, both exponentials are small,
yet the first process (late splitting) dominates over the first one, since it is less likely 
for two particles to survive than for a single one.

 \begin{figure}[t]
	\centering
	\includegraphics[width=0.7\textwidth]{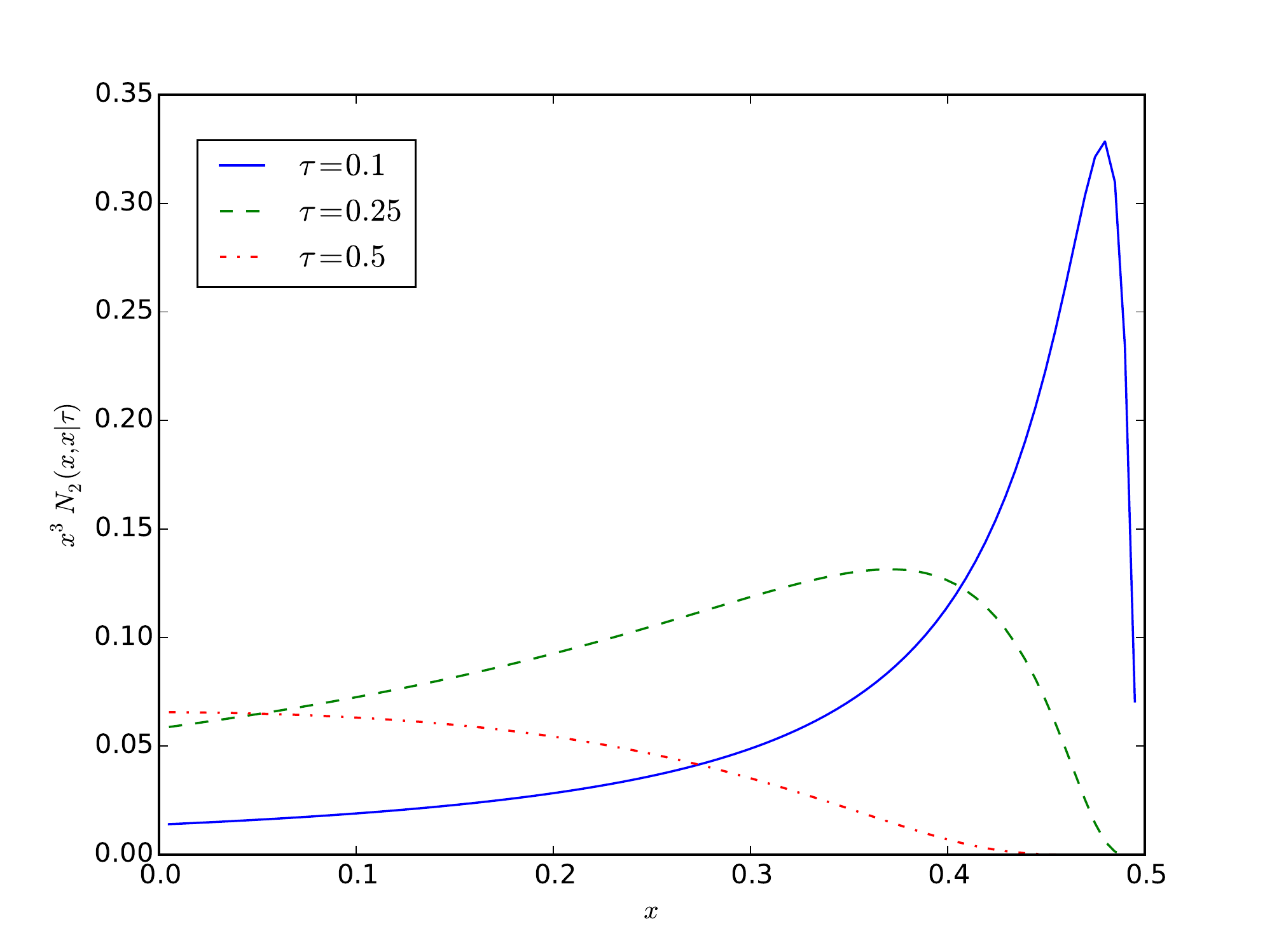}
		\caption{\sl The gluon pair density $ \mcal{N}^{(2)}(x_1,x_2|\tau)$
		with $x_1=x_2\equiv x$ is plotted as a function of $x$ for $x\le 0.5$
		 and various values of $\tau$:
		solid (blue): $\tau=0.1$; dashed (green): $\tau=0.25$;
 dotted (red): $\tau=0.5$.
		}
		\label{fig:N2}
\end{figure}

In practice though, we are more interested in the small-time regime $\tau\ll 1$, as
appropriate for the phenomenology of jets at the LHC. 
When $\pi\tau^2\ll 1$, the pair density \eqref{N2exact} develops a peak near $x_1+x_2=1$,
corresponding to the case where one of the measured gluons is the LP. 
(See the plot in Fig.~\ref{fig:N2} for an illustration.)
But the most 
interesting situation is when both $x_1$ and $x_2$ are small, $x_1,\,x_2\ll 1$,
as generally the case for radiation. In that case, \eqn{N2exact} reduces
to
\begin{align}\label{N2small}
   \mcal{N}^{(2)}(x_1,x_2|\tau)\,\simeq\,\frac{3}{2}\frac{\tau^2}{(x_1x_2)^{3/2}}
   \,\simeq\,\frac{3}{2}\, \mcal{N}(x_1,\tau) \, \mcal{N}(x_2,\tau)\,,
  \end{align}  
where we have also used the corresponding estimate for the gluon spectrum, that is,
$\mcal{N}(x,\tau)\simeq \tau/x^{3/2}$ (cf. the discussion after \eqn{Nexact}). In spite of
its factorized structure, the pair density in \eqn{N2small} does still encode strong correlations,
as shown by the following argument: the genuine 2-particle correlation is measured by the
cumulant pair density
\beq\label{C2}
\mcal{C}^{(2)}(x_1,x_2|\tau)\,\equiv\,\mcal{N}^{(2)}(x_1,x_2|\tau)-\,
\mcal{N}(x_1,\tau) \, \mcal{N}(x_2,\tau)\,\simeq\,
\frac{1}{2}\, \mcal{N}(x_1,\tau) \, \mcal{N}(x_2,\tau)\,,
\eeq
where the second estimate, valid in the regime of \eqn{N2small}, is parametrically
as large as $\mcal{N}^{(2)}(x_1,x_2|\tau)$ in that regime. As argued in \cite{Escobedo:2016jbm},
this correlation comes from processes where the LCA is itself soft ($\xi_1+\xi_2\ll 1$ in
 \eqn{N2sol}). The other possibility, namely that the LCA be the LP, does not generate 
correlations, since successive, soft emissions by the LP are nearly independent from each other.

\begin{figure}[t]
	\centering
	\includegraphics[width=0.9\textwidth]{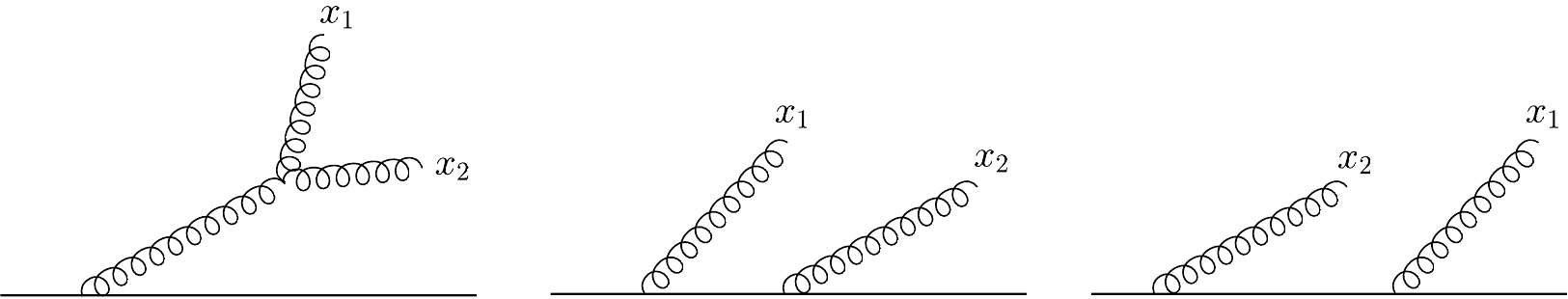}
		\caption{\sl Processes contributing to the production of 2 soft gluons
		to lowest order in perturbation theory. The process on the left, where the
		LCA  is itself soft, is the only one to generate genuine correlations.
		}
		\label{fig:2g}
\end{figure}

To leading order in perturbation theory --- by which we mean the expansion in the
number of gluon emissions (or, equivalently, the perturbative solution
to the master equations obtained via iterations) 
---, the result in \eqn{N2small} receives contributions from
the three processes shown in Fig.~\ref{fig:2g}, each of them involving two soft emissions.
It is quite easy to explicitly compute the respective contributions
(this requires two iterations of Eqs.~\eqref{eqN} and \eqref{eqN2}) and thus check 
that they sum up to the result shown in \eqn{N2small}; each channel contributes 
$1/3$ of the total result and the net correlation comes from the leftmost channel, 
where the LCA is itself soft.
However, in reality, there are arbitrarily many other processes, involving the emissions
of unresolved gluons with energies $\omega\lesssim \obr(t)$ (or
energy fractions $x\lesssim \tau^2$), which contribute to the same accuracy.
Indeed, as discussed after \eqn{omegabr}, the probability for such a soft emission
is of order one, so one can include arbitrarily many of them without modifying 
the perturbative accuracy of a calculation. The fact that such additional emissions
do not modify the net result beyond the `naive' leading-order calculation is again
a consequence of wave turbulence --- that is, of the
precise cancellation between `gain' and `loss' contributions to the r.h.s. of \eqn{eqN2}.

\subsection{The 3-gluon correlation}
\label{sec:3point}

In the Appendix \ref{sec:appNN}, we shall construct an inductive argument allowing one to compute the
factorial moments $\mcal{N}^{(p)}$ for arbitrary $p$. Before discussing the general case,
in the next subsection, let us here present and
discuss the respective result for $p=3$. This is conveniently written as
\begin{align}\label{N3exact}
\mcal{N}^{(3)}(x_1,x_2,x_3|\tau)=&\frac{3}{4\pi(x_1x_2x_3)^{3/2}}\bigg[
\frac{1}{2}\,\text{erfc}\left(\frac{\tau\sqrt{\pi}}{\sqrt{1-x_1-x_2-x_3}}\right)
\nonumber\\
&\qquad
-\text{erfc}\left(\frac{2\tau\sqrt{\pi}}{\sqrt{1-x_1-x_2-x_3}}\right)
+\frac{1}{2}\,\text{erfc}\left(\frac{3\tau\sqrt{\pi}}{\sqrt{1-x_1-x_2-x_3}}\right)\bigg]
\end{align}
where we have introduced the complimentary error function,
\beq\label{eq:erfc}
\text{erfc}(a)\equiv 1- \text{erf}(a)\,= \frac{2}{\sqrt{\pi }}\int _a^\infty\rmd z \,\rme^{-z^2}
\,=\,\frac{\rme^{-a^2}}{a\sqrt{\pi }}\left[1 +\order{\frac{1}{a^2}}\right].
\eeq
The expansion of the error function for small values of its argument will be useful too
for what follows:
\beq\label{eq:erf}
\text{erf}(a)\equiv \frac{2}{\sqrt{\pi }}\int _0^a\rmd z \,\rme^{-z^2}\,=\,
\frac{2}{\sqrt{\pi }}\left(a -\frac{a^3}{3} + \frac{a^5}{10} +\order{a^7}\right) \,.
\eeq

The special linear combination of error functions appearing in \eqn{N3exact}
has been generated via the following double time integral (compare
to \eqn{int2p})
\beq\label{int3p}
\int_0^\tau\rmd\tau_2\int_0^{\tau_2}\rmd\tau_1
(3\tau-\tau_2-\tau_1)\,\rme^{-\frac{\pi(3\tau-\tau_2-\tau_1)^2}{1-x_1-x_2-x_3}}
=\frac{1-x_1-x_2-x_3}{2\pi}\int_0^\tau \rmd\tau_2\int_0^{\tau_2}\rmd\tau_1\,
\frac{\rmd}{\rmd \tau_1}\,\rme^{-\frac{\pi(3\tau-\tau_2-\tau_1)^2}{1-x_1-x_2-x_3}}
\,.\eeq
In this integral $\tau_1$ represents the splitting time for the branching generating the
2-point correlation $\mcal{N}^{(2)}$, which subsequently acts as a source for the
3-point function (cf. \eqn{eqNp} with $p=3$). Furthermore, $\tau_2$ refers to the branching 
which creates the final 3-point correlation; that is, this is the same as the integration
variable $\tau'$ in \eqn{Npsol}. As emphasized in \eqn{int3p}, the integrand can be 
written as a total derivative w.r.t. $\tau_1$; hence the integral over $\tau_1$ is trivial and 
the subsequent integral over $\tau_2$ is recognized as the definition of the error function.

The 3 terms within the square brackets in 
\eqn{N3exact} correspond to the 3 possible combinations of late and early emissions.
The first term, which yields the dominant contribution at large times $\pi\tau^2\gtrsim 1$,
represents processes where both splittings occur very late, close to the time of measurement:
$\tau_1\simeq\tau_2\simeq\tau$. Accordingly,  this term is proportional to the survival
probability for a common ancestor with energy fraction $x_1+x_2+x_3$. [This becomes obvious after
using the asymptotic behavior of $\text{erfc}(a)$ at large $a\gg 1$, cf. \eqn{eq:erfc}.] The last
error function in \eqn{N3exact} corresponds to the case where both emissions occur very early,
$\tau_1\simeq\tau_2\simeq 0$, while the intermediate one, to the case where $\tau_1\simeq 0$
and $\tau_2\simeq \tau$ (an early emission plus a late one). These last 2 terms are strongly suppressed
at late times, since proportional with the survival probabilities for systems of 3 and, respectively, 2 particles.

\begin{figure}[t]
	\centering
	\includegraphics[width=0.9\textwidth]{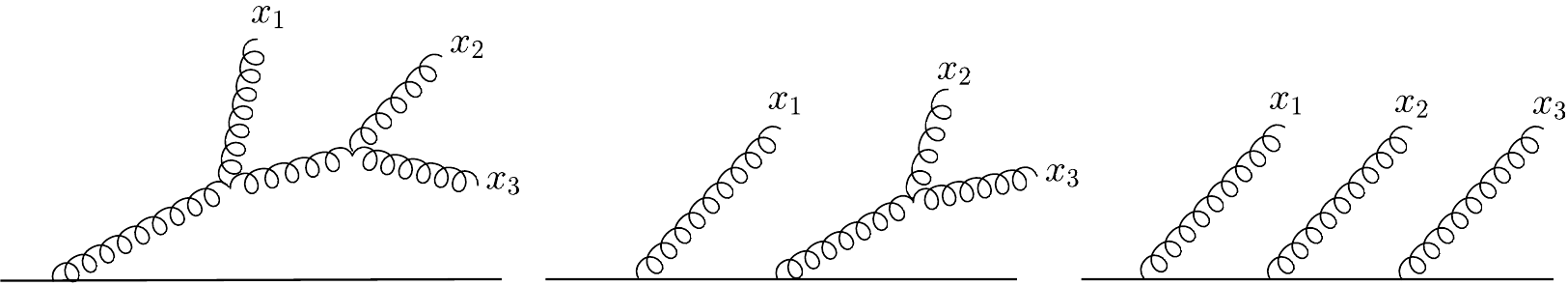}
		\caption{\sl Processes contributing to the production of 3 soft gluons
		to lowest order in perturbation theory. The process on the left, where the
		LCA  is itself soft, is the only one to generate a 3-particle correlation.
		}
		\label{fig:3g}
\end{figure}

As already explained, the most interesting situation for the phenomenology at the LHC
is the small-$\tau$ regime at $\pi\tau^2\ll 1$. In this regime and for small energy fractions\footnote{The
behavior near the kinematical limit at $x_1+x_2+x_3=1$, where the argument 
$\tau/\sqrt{1-x_1-x_2-x_3}$ of the error functions can be large even for small values of $\tau$,
is not that interesting since the 3-point function $\mcal{N}^{(3)}$ is strongly suppressed in that
limit, as we shall see in Sect.~\ref{sec:ppoint}.}
$x_i\ll 1$
(corresponding to the bulk of the radiation), one can use the expansion of the error function
in \eqn{eq:erf} to find 
\begin{align}\label{N3small}
   \mcal{N}^{(3)}(x_1,x_2,x_3|\tau)\,\simeq\,\frac{3\tau^3}{(x_1x_2 x_3)^{3/2}}
   \,\simeq\,{3}\, \mcal{N}(x_1,\tau) \, \mcal{N}(x_2,\tau)\, \mcal{N}(x_3,\tau)\,.
  \end{align}  
Note that the would-be dominant terms, linear in $\tau$, have cancelled out between the various error
functions, hence the dominant contribution is cubic in $\tau$, as it should in order to be consistent
with factorization. In spite of this factorized structure, the small-time result in \eqn{N3small}
does still encodes strong
correlations, as obvious when computing the respective cumulant (cf. \eqn{C23}):
\beq\label{C3}
\mcal{C}^{(3)}(x_1,x_2,x_3|\tau)\,\simeq\,\frac{1}{2}\,\frac{\tau^3}{(x_1x_2 x_3)^{3/2}}
\,\simeq\,\frac{1}{2}\, \mcal{N}(x_1,\tau) \, \mcal{N}(x_2,\tau)\, \mcal{N}(x_3,\tau)
\,.\eeq
To lowest order in perturbation theory, the result in \eqn{N3small} receives contributions
from processes involving the emission of exactly 3 soft gluons, as illustrated in Fig.~\ref{fig:3g}.
In particular, the net correlation in \eqn{C3} is generated by the leftmost process, where the
3 measured gluons have a soft common ancestor. As already discussed in relation with the
2-point function, this result \eqref{N3small} is truly non-perturbative, in that it receives 
contributions from processes with arbitrarily many unresolved soft gluons. However,
all contributions beyond order $\tau^3$ exactly cancel because of the fine cancellations 
between gain and loss terms --- i.e. because of wave turbulence.

\subsection{The generic $p$-body density}
\label{sec:ppoint}

The previous discussions of the gluon pair density $ \mcal{N}^{(2)}$, in Sect.~\ref{sec:pair},
and of the 3-body density $ \mcal{N}^{(3)}$, in Sect.~\ref{sec:3point}, were quite similar to
each other and this similarity has inspired us an induction argument which allows for the calculation of the 
higher-point correlations. This argument will be presented in detail in the Appendix \ref{sec:appNN}. Here, we shall
merely present the final result and discuss some physical consequences.

Specifically, our result for $ \mcal{N}^{(p)}$ can be conveniently written as 
\beq\label{Npexact}
\mcal{N}^{(p)}(x_1,\cdots,x_p|\tau)=\frac{(p!)^2}{2^{p-1}p}\frac{(1-\sum_{i=1}^p x_i)^{\frac{p-3}{2}}}{(x_1\cdots x_p)^{3/2}}h_p\left(\frac{\tau}{\sqrt{1-\sum_{j=1}^px_j}}\right)\,,
\end{equation}
with the function $h_p(\ell)$ defined by the following multiple integral,
\beq\label{hp}
h_p(\ell)=\int_0^\ell\,\rmd\ell_{p-1}\cdots\int_0^{\ell_2}\,\rmd\ell_1\Big(p\ell-\sum_{i=1}^{p-1}\ell_i\Big)\,
\rme^{-\pi\big(p\ell-\sum_{j=1}^{p-1}\ell_j\big)^2}\,,
\eeq
which is recognized as a generalization of the previous integrals appearing in the calculation
of the 2-point and 3-point functions, cf. Eqs.~\eqref{int2p} and \eqref{int3p}.
In fact, \eqn{Npexact} also covers the case of the 1-point function, i.e., the gluon spectrum 
$\mcal{N}(x,\tau)$: indeed, for $p=1$, 
we are only left with the {\em integrand} in \eqn{hp}, that is,
$h_1(\ell)=\ell\, \rme^{-\pi\ell^2}$ with $\ell=\tau/\sqrt{1-x}$; then, \eqn{Npexact}
with $p=1$ reduces indeed to \eqn{Nexact} for the gluon spectrum. The physical meaning
of the integration variables $\ell_i$ should be quite clear by now: up to a rescaling with the
common factor $1/\sqrt{1-\sum_{j=1}^px_j}$, these are the splitting times for the successive 
branchings which create the correlations.

The time-dependence of $\mcal{N}^{(p)}$ is fully encoded in the function $h_p$ and hence
it enters only via the scaling variable $\ell\equiv {\tau}/{\sqrt{1-\sum_{j=1}^px_j}}$. 
Accordingly, the structure of the factorial moment $\mcal{N}^{(p)}$  is remarkably simple:
this is essentially the product of
the power-like spectrum produced by wave turbulence (meaning one factor of $1/x_i^{3/2}$ for
each external leg), which controls the gluon distribution at small $x$ ($x_i\ll 1$),
times a scaling function which describes the time-dependence of the multi-gluon
correlation and also its behavior near the kinematical limit at $\sum_{j=1}^px_j= 1$. 

We did not attempt to analytically perform the time integrations in  \eqn{hp} for generic 
values of $p$ (the first such integral, over $\ell_1$, is of course trivial since
the integrand in \eqn{hp} is a total derivative). But, clearly, this multiple integral is well suited
for numerical calculations and also for analytic approximations, as we shall now discuss.

We first consider the situation where the scaling variable $\ell$ is large, $\ell\gg 1$. This
includes the large-time regime, $\tau\gg 1$, but also the behavior near the kinematical limit 
at $\sum_{j=1}^px_j= 1$ for generic values of $\tau$. As already seen on the examples of
the 2-point and 3-point functions and it is also intuitive by inspection of \eqn{hp}, the dominant
behavior in this limit comes from processes where all the relevant splittings 
occur as late as possible: $\ell_i\simeq \ell$ for any $i=1,2,\dots, p-1$. Indeed, such
configurations minimize the exponent of the Gaussian within the integrand of \eqn{hp}.
We thus expect an asymptotic behavior $h_p(\ell)\propto \rme^{-\pi\ell^2}$, proportional
to the survival probability of the last common ancestor (with energy fraction $x_1+x_2 +
\cdots +x_p$) over a time of order $\tau$. This is confirmed by the manipulations in the
Appendix \ref{sec:largel}, which more precisely yield
\begin{equation}
\label{hplarge}
h_p(\ell)\,\simeq\,\frac{\rme^{-\pi \ell^2}}{\ell^{p-2}(2\pi)^{p-1}(p-1)!}\qquad\mbox{for}\quad \pi\ell^2 \gg 1\,.
\end{equation}
This holds up to corrections suppressed by inverse powers of $\ell$ and/or exponentials
factors like $\rme^{-4\pi \ell^2}$. This approximation yields
\beq\label{Npasymp}
\mcal{N}^{(p)}(x_1,\cdots,x_p|\tau)\,\simeq\,\frac{p!}{(4\pi)^{p-1}\tau^{p-2}}
\,\frac{(1-\sum_{i=1}^p x_i)^{n-5/2}}
{(x_1\cdots x_p)^{3/2}}\,\exp\left\{-\frac{\pi\tau^2}{\sqrt{1-\sum_{j=1}^px_j}}\right\}
\,.
\end{equation}
This is in agreement with our previous results  for the
$p$-point functions with $p=1,2,3$. Notice that, unlike the
gluon spectrum \eqref{Nexact} and the pair density  \eqref{N2exact}, which exhibit a 
leading-particle peak in the vicinity of the kinematical limit,
the higher-point correlations with $p\ge 3$ do not show such a peak --- rather,
they rapidly vanish when approaching that limit.

We now turn to the more interesting regime at small times $\tau\ll 1$ and soft produced
particles $x_i\ll 1$. Clearly, this means $\ell\simeq\tau \ll 1$. It is then tempting to evaluate the
integrations in \eqn{hp} by using the small-argument expansion of the exponential.
For this to be justified for generic values of $p$ (including larger values $p\gg 1$), 
one however needs the stronger condition $\pi (p\ell)^2\ll 1$. Under this stronger
assumption, the dominant behavior is indeed obtained by replacing the Gaussian by unity. One thus
finds (see the Appendix \ref{sec:smalll} for details)
\beq\label{hpsmall}
h_p(\ell)\simeq \int_0^\ell\,\rmd\ell_{p-1}\cdots\int_0^{\ell_2}\,\rmd\ell_1\Big(p\ell-\sum_{i=1}^{p-1}\ell_i\Big)
=\,\frac{(p+1)\ell^{\,p}}{2(p-1)!}\qquad\mbox{for}\quad \pi (p\ell)^2 \ll 1
\,,
\eeq
and therefore
\beq\label{Npsmall}
\mcal{N}^{(p)}(x_1,\cdots,x_p|\tau)\simeq \frac{(p+1)!}{2^{p}}\,\frac{\tau^p}
{(x_1\cdots x_p)^{3/2}}\qquad\mbox{for}\quad \pi \tau^2\ll \frac{1}{p^2}\quad
\mbox{and} \quad x_i\ll 1
\,,\eeq
in agreement with our respective results for $p=1,2,3$. This power-like multi-particle spectrum,
with the characteristic exponent $3/2$, is the consequence of wave turbulence for the jet problem at hand.
In spite of the factorized structure of \eqn{Npsmall}, this result still encodes genuine correlations, as already
discussed on the examples of the 2-point and 3-point functions. These correlations are generated
by processes where all the $p$ measured particles have
a soft common ancestor --- a primary gluon whose evolution via democratic branchings has generated
the mini-jet to which all the measured particles belong. As explained, the genuine correlations can be 
isolated by computing the factorial cumulant. As an example,
we here show the corresponding result\footnote{One
has (with compact notations, whose meaning should be obvious):
 $\mcal{C}^{(4)}_{1234}=\mcal{N}^{(4)}_{1234}  - \big[\mcal{N}^{(2)}_{12}\mcal{N}^{(2)}_{34}
+ 2\ \mbox{perms.}\big] - \big[\mcal{N}_1\mcal{N}^{(3)}_{234} + 3\ \mbox{perms.}\big]
+2\big[\mcal{N}_1 \mcal{N}_2 \mcal{N}^{(2)}_{34}
+ 5\ \mbox{perms.}\big] -6\mcal{N}_1\mcal{N}_2\mcal{N}_3\mcal{N}_4$.}
for $p=4$~: 
\beq\label{C4small}
\mcal{C}^{(4)}(x_1,x_2,x_3,x_4|\tau)\simeq \frac{3}{4}\,\frac{\tau^4}
{(x_1 x_2 x_3 x_4)^{3/2}}
\,.\eeq

 \begin{figure}[t]
	\centering
	\includegraphics[width=0.7\textwidth]{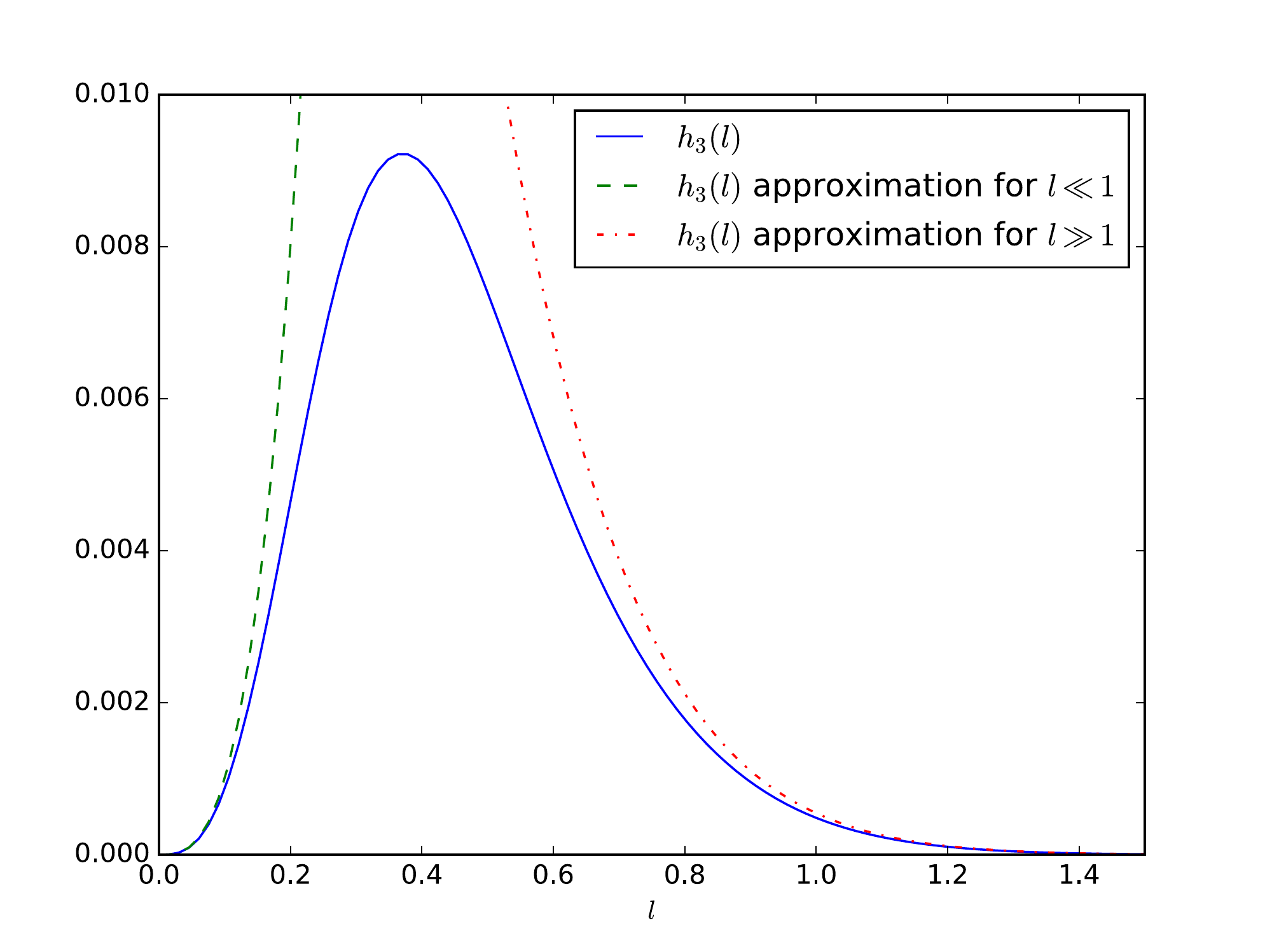}
		\caption{\sl The scaling function $h_3(\ell)$,  computed
		according to \eqn{hp} [this can be also read from \eqn{N3exact}], is compared to its
		approximations at $\ell\gg 1$, cf. \eqn{hplarge},
and at $\ell\ll 1$, cf. \eqn{hpsmall}.  }		
		\label{fig:h3}
\end{figure}

To illustrate the previous approximations for the function $h_p(\ell)$, we display in Fig.~\ref{fig:h3}
the function $h_3(\ell)$ (whose explicit form can be easily inferred by comparing Eqs.~\eqref{N3exact}
and \eqref{Npexact}) together with its approximate versions at $\ell\gg 1$, cf. \eqn{hplarge},
and at $\ell\ll 1$, cf. \eqn{hpsmall}.

\section{Gluon multiplicities and KNO scaling}
\label{sec:mult}

In the previous section, we have studied the detailed energy distribution of the 
medium-induced radiation, as characterized by the factorial moment 
densities $\mcal{N}^{(p)}(x_1,\cdots,x_p|\tau)$.
In what follows, we shall `integrate out' the distribution in energy in order to deduce the statistics
of the {\em gluon multiplicities} --- the total number of gluons and its fluctuations. 

\subsection{Multiplicities for soft gluons}
\label{sec:number}

If one attempts to compute the average number of gluons by integrating the
gluon spectrum \eqref{Nexact} over $x$, that is, $\langle N(\tau)\rangle 
=\int_0^1\rmd x \,\mcal{N}(x,\tau)$,
then one faces a severe infrared ($x\to 0$) divergence, due to the strong, power-like, 
enhancement in the gluon density at small $x$ :  $\mcal{N}(x,\tau)\propto 1/x^{3/2}$.
This argument shows that the total gluon number is not a meaningful observable, since the
radiation produces infinitely many soft gluons. Rather, it makes sense to compute the total
number of gluons with energies larger than some minimal value (`infrared cutoff') $\omega_0$,
meaning with energy fractions $x\ge x_0$, where $x_0\!\equiv\! \omega_0/E$. (We recall that $E$
is the initial energy of the leading particle.) A natural value for $\omega_0$ exists on
physical grounds: this is the characteristic energy of the medium, say, its temperature $T$ if
the medium is a weakly-coupled quark-gluon plasma in (or near) thermal equilibrium.

Indeed, the ideal branching dynamics considered so far strictly applies only so long as
the energies of the gluons from the cascade remain much larger than this medium scale,
$\omega\gg T$. On the other hand, when $\omega\sim T$, the dynamics is modified, first,
by the elastic collisions between the gluons from the jet and the constituents of the medium
and, second, by the non-linear effects associated with the relatively large gluon occupation
numbers (which for $\omega\sim T$ become of $\order{1}$, since one cannot distinguish
anymore between gluons from the jets and those from the medium). As a result of such
modifications, the soft gluons with $\omega\lesssim T$ are expected to thermalize, which
in turn will stop the branching process (due to the detailed balance between
splitting and recombination processes) \cite{Iancu:2015uja}.

For the kinematical conditions at the LHC, one has $T\ll \obr(L)\ll E$, where we recall
that $\obr(L)=\abar^2\hat q L^2$ is the characteristic energy scale for the onset of multiple branching
(cf. the discussion after \eqn{omegabr}). The first inequality 
ensures that one has a sufficiently large phase-space at soft momenta for the
jet evolution via multiple branchings to be fully developed. The second  inequality implies
that the medium is relatively thin, $L\ll \tbr(E)$, so the leading particle survives in the
final state --- it radiates soft gluons with $\omega\lesssim \obr(L)$, but it
cannot undergo a democratic branching. 

In practice, it is useful to chose the infrared cutoff $\omega_0$ to be much smaller than 
$\obr(L)$ (in order to probe the physics of multiple branchings) but still larger
than $T$ (to be able to distinguish the particles from the jet from those in the surrounding
medium).  Accordingly, we shall compute the $p$-order factorial moment of the multiplicity 
as follows,
\begin{align}\label{avNpdef}
\langle N(N-1) \dots (N-p+1)\rangle (\tau,x_0)=\int_{x_0}^1\rmd x_1 \dots
\int_{x_0}^1\rmd x_p
\,\mcal{N}^{(p)}(x_1,\cdots,x_p|\tau)\,.
\end{align}
with the lower cutoff $x_0$ satisfying $T/E < x_0 \ll \tau^2 \lesssim 1$. The
result is strongly sensitive to the precise value of $x_0$, yet in what follows we shall
identify observables which are independent of this cutoff.
For the physical interpretation of the subsequent results, it is useful to recall the
relations
\beq\label{tau}
\tau^2\,= \,\frac{L^2}{\tbr^2(E)}\,=
\,\frac{\abar^2 \hat q L^2}{E}\,=\,\frac{\obr(L)}{E}\,.
\eeq

 \begin{figure}[t]
	\centering
	\includegraphics[width=0.7\textwidth]{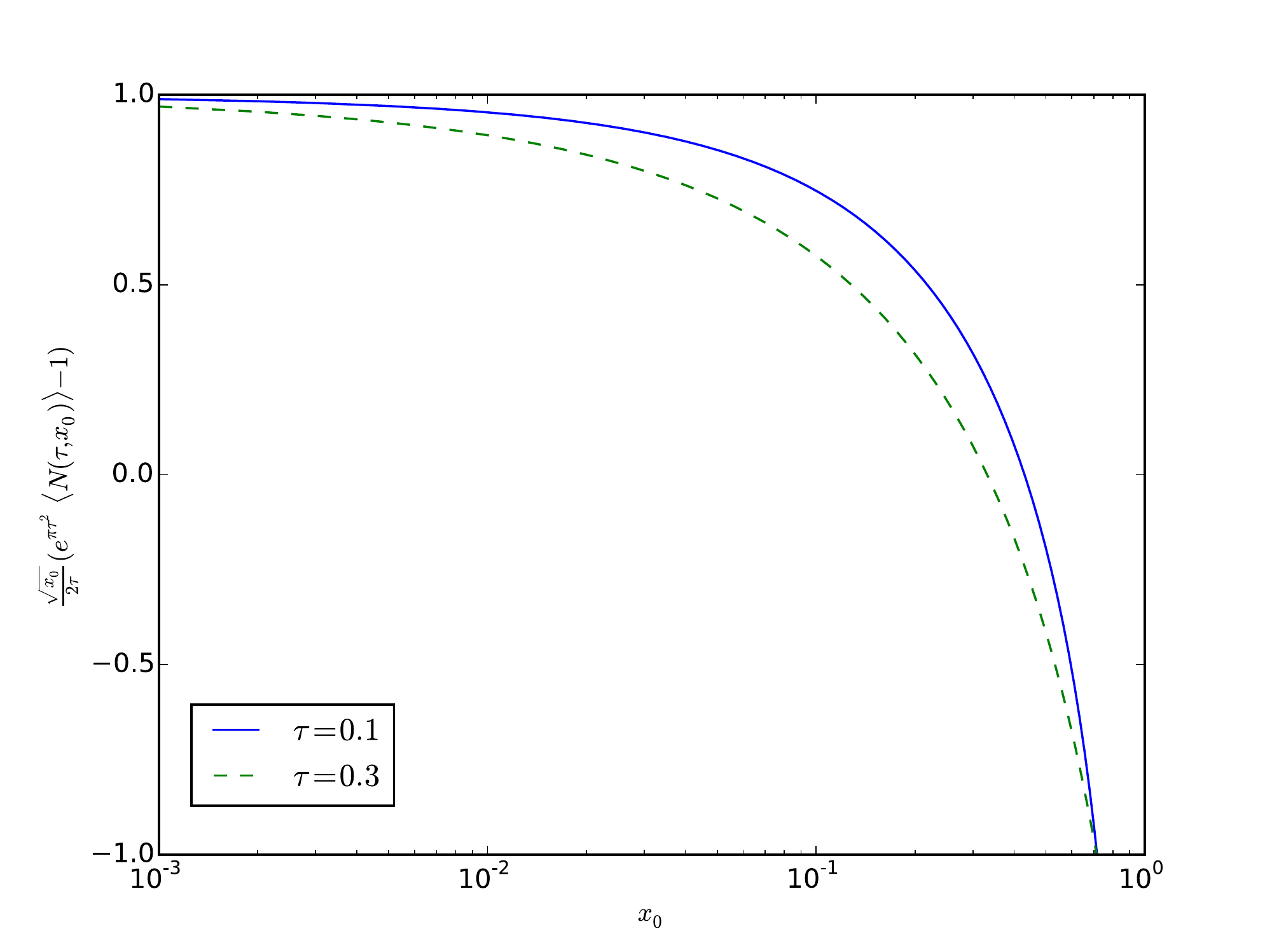}
		\caption{\sl The ratio between the exact result \cite{Escobedo:2016jbm}
		$\langle N\rangle (\tau,x_0)-\rme^{-\pi\tau^2}$
		for the average multiplicity minus the leading particle and the small-$x_0$ approximation
		to  $\langle N\rangle (\tau,x_0)$ shown in  \eqn{avNgeom} is plotted as a function
		of $x_0$ for two values of $\tau$.}		
		\label{fig:Nlog}
\end{figure}

When $x_0\ll\tau^2$, i.e. $\omega_0\ll \obr(L)$, the multiple integral in \eqn{avNpdef} is controlled
by its lower limit, i.e. $x_i\sim x_0$ for any $i=1,\dots,p$, due to the copious production
of soft gluons via multiple branching. One can therefore estimate this integral by using the
approximate version of $\mcal{N}^{(p)}(x_1,\cdots,x_p|\tau)$ valid when $x_1+x_2+\dots+x_p\ll 1$,
that is (cf. \eqn{Npexact}),
\beq\label{Npsmallx}
\mcal{N}^{(p)}(x_1,\cdots,x_p|\tau)\simeq
\frac{(p!)^2}{2^{p-1}p}\frac{1}{(x_1\cdots x_p)^{3/2}}\,h_p\left(\tau\right)
\qquad\mbox{for}\quad \sum_i x_i\ll 1\,.
\end{equation}
Then the integrals in \eqn{avNpdef} become trivial and yield
\beq\label{avNgeom}
\langle N(N-1) \dots (N-p+1)\rangle (\tau,x_0)\simeq \frac{2(p!)^2}{p}\,
\frac{h_p\left(\tau\right)}{x_0^{p/2}}\qquad\mbox{for}\quad x_0\ll \tau^2\,.\eeq
As it should be clear from the above, this approximation properly counts the soft gluons
produced via radiation, but it ignores the relatively hard gluons
with $\tau^2\lesssim x \le 1$ and notably the leading particle. Hence, this should be a good
approximation so long as the gluon multiplicities are very high, in particular $\langle N\rangle (\tau,x_0)
\gg 1$, which is strictly true so long as $x_0\ll\tau^2$. In practice though, this remains a good
approximation up to larger values $x_0\sim\tau^2$ provided one pays attention not to include
the LP when counting the multiplicity --- a condition which is easy to fulfill in the experimental situation,
where the LP can indeed be distinguished from its soft products of radiation so long as $\tau < 1$.
To illustrate this, we compare in Fig.~\ref{fig:Nlog}
the prediction of \eqn{avNgeom} for the average multiplicity $\langle N\rangle (\tau,x_0)$
against the exact respective result \cite{Escobedo:2016jbm} 
{\em from which we subtract the contribution of
the LP}. That is, \eqn{avNgeom} with $p=1$ is compared to $\langle N\rangle (\tau,x_0)-\rme^{-\pi\tau^2}$,
with $\langle N\rangle (\tau,x_0)$ given by Eq.~(4.4) in Ref.~\cite{Escobedo:2016jbm}.
As visible in this figure, the approximation in  \eqn{avNgeom}
is indeed accurate up to $x_0\sim\tau^2$.

\subsection{KNO scaling}
\label{sec:KNO}

A remarkable feature of the result in \eqn{avNgeom} is the fact that it scales as a power of $1/x_0$,
with an exponent proportional to $p$. This reflects the power-like spectrum of the factorial 
moments at small $x_i$, cf. \eqn{Npsmallx}, which we recall is a (turbulent) fixed point of the
evolution via multiple branching. In turn, this implies that the dependence upon
$x_0$ cancels out when constructing the {\em reduced moments},
\beq\label{kappapdef}
\kappa^{(p)} (\tau,x_0)
\equiv\frac{\langle N(N-1) \dots (N-p+1)\rangle}{\langle N\rangle^p}\,\simeq\,
\frac{p! (p-1)!}{2^{p-1}}\,\frac{h_p(\tau)}{\big[h_1(\tau)\big]^p}
\qquad\mbox{for}\quad x_0\ll \tau^2\,.
\eeq

The reduced moments \eqref{kappapdef}
exhibit {\em geometric scaling} :  they depend upon the physical parameters
$L$,  $\hat q$,  and $E$ only via the dimensionless variable $\tau=\abar L\sqrt{{\hat q}/{E}}$.
This scaling takes a particularly simple form at sufficiently small $\tau$, when one can
use the approximation \eqref{hpsmall} for $h_p(\tau)$. In that case, the ratio
$h_p(\tau)/[h_1(\tau)]^p\simeq (p+1)/[2(p-1)!]$ becomes independent of $\tau$, hence the
reduced moment \eqref{kappapdef}  is a {\em pure number}, which depends only upon $p$~:
\beq\label{kappapsmall}
\kappa^{(p)}(\tau, x_0)\, \simeq\,\frac{(p+1)!}{2^p}
\qquad\mbox{for}\qquad x_0\ll \tau^2\ll \frac{1}{p^2}\,.\eeq
This property is known as  {\em  KNO scaling} (from Koba, Nielsen, and Olesen \cite{Koba:1972ng}).
In a previous publication \cite{Escobedo:2016jbm}, we have obtained this result \eqref{kappapsmall} for 
the particular case $p=2$ and conjectured the emergence of KNO scaling for generic values of $p$.
The present analysis confirms the existence of this scaling, clarifies the limits of its validity (cf. the inequalities
in the r.h.s. of \eqref{kappapsmall}), and also specifies the corresponding value for
$\kappa^{(p)}$ for any $p\ge 2$. As we shall shortly explain, this value for $\kappa^{(p)}$  is 
quite special and allows us to identify the probability distribution for gluon multiplicities at small $\tau$.

 \begin{figure}[t]
	\centering
	\includegraphics[width=0.7\textwidth]{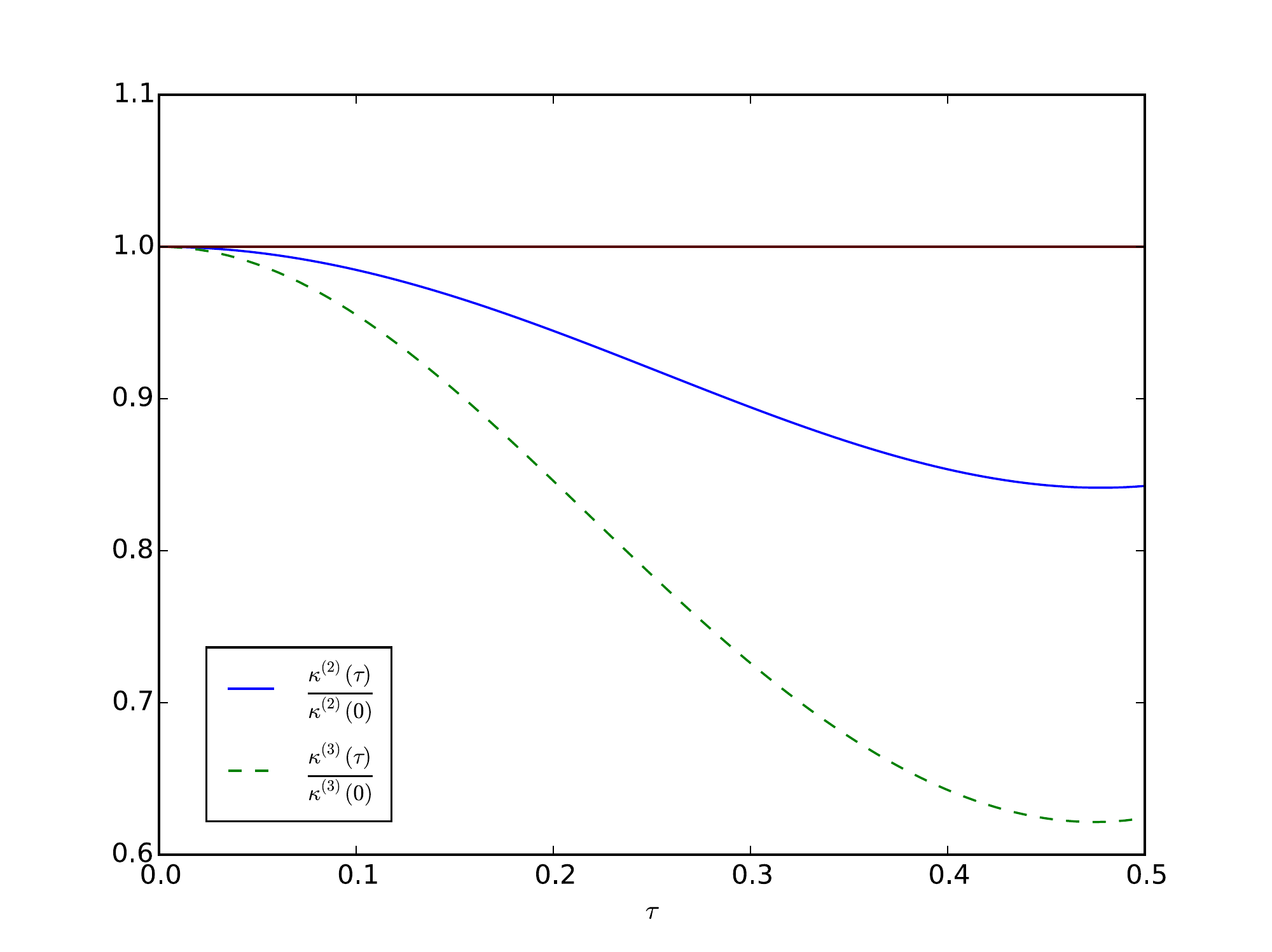}
		\caption{\sl The reduced moments $\kappa^{(p)} (\tau)$ with $p=2$ and $p=3$,
		 computed according to \eqn{kappapdef} (and conveniently normalized by 
		 the respective values at $\tau=0$), are displayed as
		 a function of $\tau$.}		
		\label{fig:kappa}
\end{figure}

In practice, the KNO scaling is restricted to rather small values of $\tau$ --- the more so,
the larger the value of $p$. This limitation is indeed visible in  Fig.~\ref{fig:kappa}, 
where we plot the ratio $\kappa^{(p)} (\tau)/\kappa^{(p)}(0)$ 
computed according to \eqn{kappapdef} as a function of $\tau$,
for $p=2$ and $p=3$. The respective KNO prediction, namely
$\kappa^{(p)} (\tau)/\kappa^{(p)} (0)=1$ (cf. \eqn{kappapsmall}), is seen to be satisfied
only at very small values of $\tau$. Besides, the deviation from it with increasing $\tau$
starts earlier (and grows faster) for $p=3$ than for $p=2$.
Physically, this can be understood as follows: the scaling occurs so long
as {\em all} the configurations (in the sense of branching trees within the parton cascade) that
contribute to the simultaneous production of a set of $p$ particles survive with a probability
of order one. This requires $\rme^{-\pi(p\tau)^2}\sim \order{1}$, hence $\tau^2\lesssim 1/(\pi p^2)$.
The KNO scaling should be better and better satisfied with increasing the jet
energy $E$ (since $\tau^2\propto 1/E$), but it is unclear 
whether this can be observed within the current experimental conditions at the LHC.

Incidentally, under the assumptions of \eqn{kappapsmall} it is possible to obtain a
relatively compact expression for the factorial moments themselves (and not only for
their ratios), namely
\beq\label{NpKNO}
\langle N(N-1) \dots (N-p+1)\rangle (L,\omega_0)\simeq \,(p+1)!\left[
\frac{\obr(L)}{\omega_0}\right]^{p/2}
\quad\mbox{for}\quad \omega_0\ll \obr(L)\ll \frac{E}{p^2}\,.\eeq
As indicated by the above notations, the soft gluon multiplicities in this high-energy regime
become independent upon the energy $E$ of the LP, but only depend upon the (large) ratio
$\obr(L)/\omega_0$ between the characteristic medium scale $\obr(L)$ and the
energy resolution scale $\omega_0$. Moreover, the multiplicities are parametrically 
large in this regime, hence the factorial moments can be identified with the ordinary  moments:
one has $\langle N(N-1) \dots (N-p+1)\rangle  \simeq \langle N^p\rangle$ for any fixed value of $p$.

 \begin{figure}[t]
	\centering
	\includegraphics[width=0.7\textwidth]{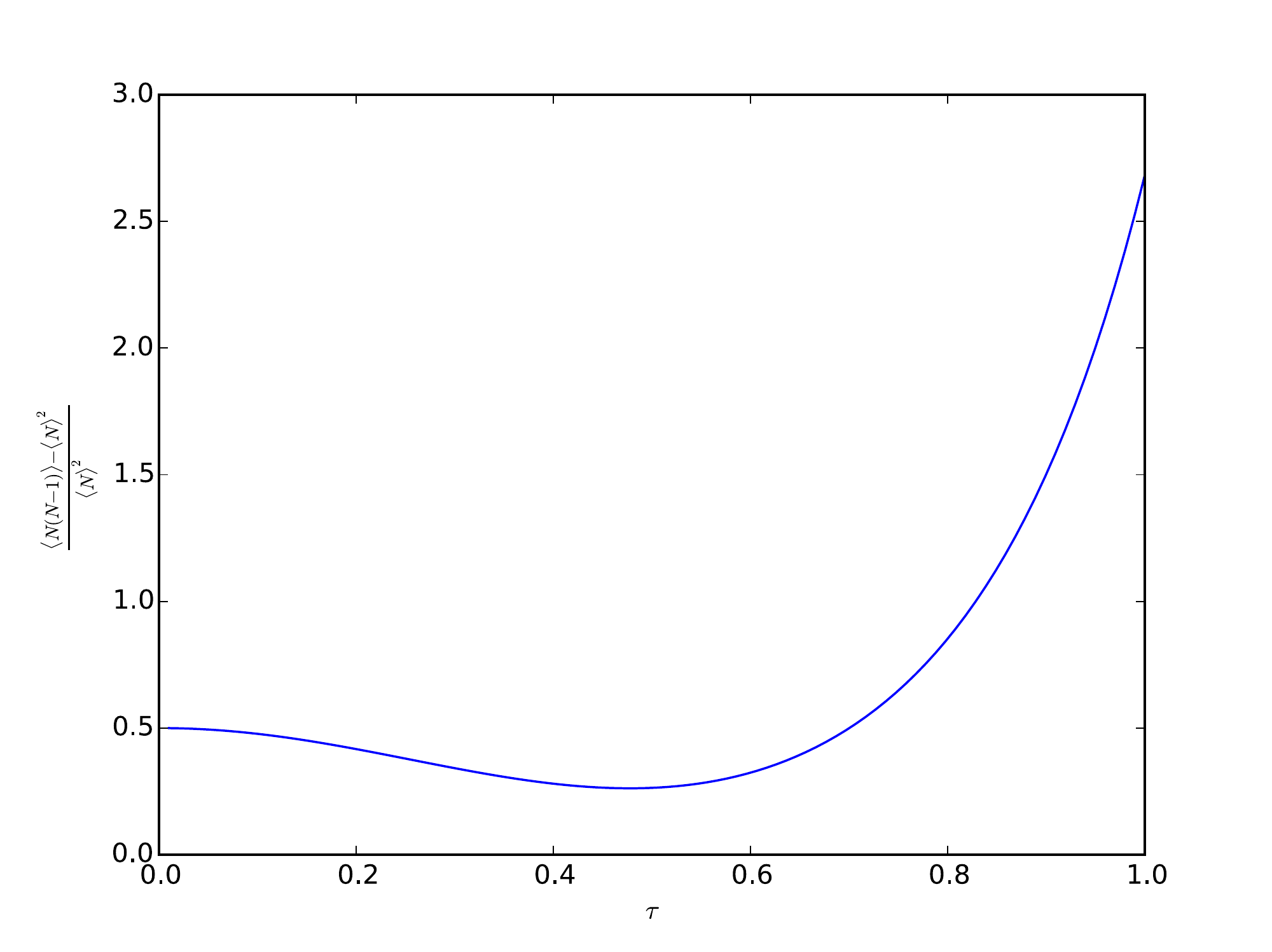}
		\caption{\sl The reduced second order cumulant computed according to
		\eqn{C2red}.}
		\label{fig:kappa2}
\end{figure}

Let us finally emphasize a rather general and also robust conclusion of the 
previous discussion, namely, the fact that the {\em fluctuations in the soft multiplicities are very large}.
This becomes more transparent if one considers the respective cumulants, like 
$\langle N(N-1) \rangle - \langle N\rangle^2$, which are a direct measure of correlations.
Recall indeed that all the cumulants vanish for the Poisson distribution, which describes
independent emissions. But for the present distribution, the cumulants are non-zero and large --- 
in fact, as large as possible: indeed, they are parametrically as large as the factorial
moments themselves. For instance, using \eqn{kappapdef} for $x_0\ll \tau^2$, one finds 
\beq\label{C2red}
\frac{\langle N(N-1) \rangle - \langle N\rangle^2}{\langle N\rangle^2}\simeq
\frac{h_2(\tau)}{\big[h_1(\tau)\big]^2}-1\simeq
    \begin{cases}
        \displaystyle{1/2} &
        \text{ for\,  $\tau \ll 1$,}
        \\*[0.2cm]
        \displaystyle{
        \frac{{\rm e}^{\pi\tau^2}}{2\pi\tau^2}}&
        \text{ for\,  $\tau \gtrsim 1$\,,}
        \\*[0.2cm]
    \end{cases}
\eeq
where we have also used the approximations \eqref{hpsmall} and \eqref{hplarge}
for small and respectively large values of $\tau$. This behavior is illustrated in 
Fig.~\ref{fig:kappa2}. The result 1/2 at small $\tau$ is an immediate consequence of
\eqn{C2} and comes from processes where the 2 measured particles belong to a same
mini-jet, i.e. they have a {\em soft} common ancestor.
For larger $\tau\gtrsim 1$, the second cumulant is even larger than the disconnected piece 
$\langle N\rangle^2$; this is a consequence of the fact
that the large-$\tau$ behavior of $\langle N(N-1) \rangle$ is controlled by special
configurations where the splitting of the last common ancestor occurs towards the end of the evolution
(cf. the discussion after \eqn{int2p}). Similar conclusions apply for the higher cumulants
with $p> 2$. They show that the correlations in the gluon distribution at small $x$
are indeed very strong. A more precise characterization of these correlations at small
$\tau$ (i.e. in the high energy limit) will be presented in the next subsection.

\subsection{A negative binomial distribution}
\label{sec:NBD}

As previously mentioned, the soft gluon multiplicities in the small-$\tau$ (or high-energy)
limit do not only exhibit KNO scaling, but they also suggest a remarkable probability distribution,
that can be directly read off \eqn{kappapsmall} : this is the {\em negative binomial (or Pascal)
distribution with parameter $r=2$}. To be more specific, let us first recall that the 
negative binomial distribution (NBD) involves 2 free parameters, 
the average particle number $\bar n\equiv \langle n\rangle$ and a positive integer\footnote{Generalizations
to real values of $r$ are also possible, but they are not useful for our present purposes.} $r$
whose meaning will be shortly explained. The associated probability law reads:
\beq\label{NBDlaw}
\mcal{P}_n(\bar n, r)\,=\,\frac{(n+r-1)!}{n! (r-1)!}\,\beta^n(1-\beta)^r\,,\qquad
\beta\,\equiv\,\frac{\bar n}{\bar n+r}\,.\eeq
One possible interpretation for the random variable $n$ which is consistent with this
distribution is as follows: $n$ is the number of failures before the occurrence of
a prescribed number $r$ of successes in a sequence of independent Bernoulli trials 
(see e.g. \cite{Feller,NBD} for details). Using \eqref{NBDlaw} one can easily compute
the associated generating functional,
\beq
Z(\bar n, r | u)\,\equiv\,\sum_{n=0}^\infty u^n\,\mcal{P}_n(\bar n, r)\,=\,{\left[
1+\frac{\bar n}{r}(1-u)\right]^{-r}}\,,
\eeq
from which it is straightforward to deduce the factorial moments:
\beq\label{NBDk}
\langle n(n-1) \dots (n-p+1)\rangle=\,\frac{\del Z}{\del u}\bigg |_{u=1}
=\,\frac{(r+p-1)!}{r!\, 2^{p-1}}\,{\bar n}^p\,.
\eeq
For $r=2$, this is in agreement with \eqn{kappapsmall}, as anticipated.

We are not aware of any fundamental physical reason for the emergence of this
particular NBD in the jet problem at hand. Moreover, it should be clear from the above
that the actual multiplicity distribution inside the jet is generally different from (and more complicated
than) a NBD: in general, the reduced moments \eqref{kappapdef} depend upon an additional
parameter $\tau$, which is real and positive. They reduce to the simple form in \eqn{kappapsmall}
only for sufficiently small values of $\tau$, where what we mean by `sufficiently small' depends
upon $p$ --- the precise condition becomes more and more restrictive with increasing $p$.
This means that, strictly speaking, there is no fixed value of $\tau$, even if arbitrarily small,
for which the actual multiplicity distribution is truly equivalent to the NBD with $r=2$.

Moreover, even in the physical regime where our approximations \eqref{kappapsmall} and
\eqref{NpKNO} make sense, they do not probe the {\em details} of the NBD 
for generic values of $n$ and $\bar n$, but only its {\em tail} at large multiplicities, 
$n\gg 1$ and $\bar n\gg 1$. Indeed, \eqn{NpKNO} applies only in the high-multiplicity
(and high-energy) regime at $\omega_0\ll \obr(L)\ll {E}/{p^2}$. Vice-versa, the
expression \eqref{kappapsmall} for the reduced moment $\kappa^{(p)}$ is also generated
by the simplified version of \eqn{NBDlaw} (with $r=2$) valid at large $n$ and large $\bar n$, 
that is,
\begin{align}\label{KNO}
\bar n \mcal{P}_n(\bar n, 2)\,\simeq\, 
4\rho\,\rme^{-2\rho}\,,\qquad \rho\equiv \frac{n}{\bar n}\,.
\end{align}
Notice that, in this approximation, the quantity $\bar n \mcal{P}_n$ 
scales as a function of $\rho={n}/{\bar n}$, a property which is sometimes used
as the {\em definition} of KNO scaling.  Let us rapidly check that \eqref{KNO}
implies indeed the result \eqref{kappapsmall} for $\kappa^{(p)}$ in the 
limit where $n\gg 1$. In this limit, one can ignore the difference between
factorial and ordinary moments and replace the sum over $n$ by an integral:
\begin{align}\label{np}
\langle n^p\rangle\equiv \sum_{n=0}^\infty n^p \,\mcal{P}_n(\bar n, 2)
\,\simeq\,\int_0^\infty \frac{\rmd n}{\bar n}\,n^p\big[\bar n \mcal{P}_n\big]
\,=\,4\bar n^p \int_0^\infty \rmd \rho\,\rho^{p+1}\,\rme^{-2\rho}\,=\,
\frac{(p+1)!}{2^p}\,\bar n^p\,.
\end{align}

But even if limited, this relation between the distribution of soft gluons within the
in-medium jet and the NBD is conceptually interesting, as we now explain. 
This becomes clearer when the present problem is compared to the evolution of a jet 
in the vacuum, as driven by its virtuality. 
In that case too, one found that the soft particle multiplicities obey KNO scaling, with 
reduced moments $\kappa^{(p)}$ that have been explicitly computed 
(within a double logarithmic approximation) \cite{Dokshitzer:1991wu}. 
In particular, one found $\kappa^{(2)}=4/3$,
which via \eqn{NBDk} appears to be consistent with a NBD with parameter $r=3$.
This identification is not fully right --- for $p\ge 3$, the respective values 
$\kappa^{(p)}$ start deviating\footnote{In particular, the exponential decay of the scaling 
function $f(\rho)\equiv \bar n \mcal{P}_n(\bar n)$ at large values of $\rho\equiv {n}/{\bar n}$ 
is found as $f(\rho)\propto\rme^{-\beta_0 \rho}$ with $\beta_0\simeq 2.552$ 
\cite{Dokshitzer:1991wu}. This is different than the corresponding 
prediction of the NBD with parameter $r=3$, namely $f(\rho)\propto\rme^{-3\rho}$.}
from those of the NBD with $r=3$ ---, yet this is
representative for the statistical properties of the in-vacuum jet evolution. 
The smaller the value of $r$, the broader is a negative binomial distribution\footnote{This
is clear e.g. by inspection of the second order cumulant: 
$\langle n(n-1)\rangle  - \bar n^2= \bar n^2/r$.}. Hence,
also by this argument, we conclude that the multiplicity distribution created by the 
medium-induced evolution is considerably broader --- in the sense of developing
larger statistical fluctuations --- than that associated with a jet which propagates in the vacuum.

\section{Conclusions}
\label{sec:conc}

In this paper we have investigated the multi-particle correlations in the gluon distribution
generated via medium-induced multiple branchings by an energetic jet propagating through a weakly
coupled quark-gluon plasma. We have demonstrated that, under suitable approximations, the
jet evolution can be described as a stochastic branching process which is exactly solvable:
for this process, we have obtained exact analytic results for all the $p$-body gluon densities
in the space of energy. The corresponding results for $p=1$ (the gluon spectrum) 
\cite{Blaizot:2013hx} and $p=2$ (the gluon pair density) \cite{Escobedo:2016jbm}
were already known in the literature, but those for the higher-point correlations with $p\ge 3$ are new.
By integrating these densities over the energies of the gluons, above an infrared cutoff
$\omega_0$ which plays the role of the resolution scale, we have deduced the factorial
moments $\langle N(N-1) \dots (N-p+1)\rangle  (L,\omega_0)$ 
which characterize the distribution of the gluon multiplicity. 

The results that we have thus obtained have interesting physical consequences, 
which could be observed in the experiments. 
They demonstrate large multiplicities for the soft gluons together with strong correlations 
associated with the existence of common ancestors. 
While such correlations were to be expected in the context of a branching process,
they appear to be significantly stronger than for the corresponding process in the vacuum (the
DGLAP evolution of a jet driven by the virtuality of the leading particle) \cite{Dokshitzer:1991wu}.
This reflects the fundamental difference between the respective branching laws:
unlike the rate for bremsstrahlung in the vacuum, which is scale invariant and favors soft
splittings (i.e. splittings where the daughter gluon carries only a small fraction of the energy
of its parent parton), the BDMPSZ rate for medium-induced gluon branching involves the
dimensionful transport coefficient $\hat q$ and favors democratic splittings (at least for the
sufficiently soft gluons --- those whose energies are softer than the characteristic medium
scale $\obr=\abar^2\hat q L^2$). As a result, the medium-induced branchings are strongly 
biased towards soft energies. This leads to the abundant production of soft gluons with
$\omega\lesssim \obr$. Moreover, any such a gluon becomes the seed of a `mini-jet'
produced via a sequence of democratic branchings. All the gluons within the same mini-jet
are correlated with each other, as they have a common ancestor. 

The expressions for the soft gluon multiplicities and the associated correlations become
particularly suggestive in the high energy limit $E\gg\obr$\,: they are independent 
of the energy $E$ of the leading particle and scale as powers of the large ratio 
$\obr/\omega_0$; specifically,
$\langle N^p\rangle\propto\big[\obr/\omega_0\big]^{p/2}$. This in particular implies
that $\langle N^p\rangle$ grows with the medium size as $L^p$. It furthermore implies
that the reduced moments $\kappa^{(p)}\equiv \langle N^p\rangle/\langle N\rangle^p$
are pure numbers, independent of any of the physical parameters of the problem. This
property is known as KNO scaling. The specific value of $\kappa^{(p)}$, cf.  \eqn{kappapsmall},
is instructive too: it implies that the associated probability distribution is a special negative
binomial distribution, which is {\em over-dispersed} --- i.e., it features large fluctuations.
This appears to be more dispersed than the distribution produced by
the DGLAP evolution of a jet propagating in the vacuum.

It would be interesting to search for confirmations of these results in the experimental
data at the LHC --- notably, in the distribution of soft particles at large angles in the 
context of di-jet asymmetry. Most likely, it should be difficult to see indications
of the KNO scaling: on one hand, our prediction in that sense relies on idealized
theoretical assumptions; on the other hand, it is notoriously difficult to experimentally
measure multi-particle correlations in nucleus-nucleus collisions, due to the large 
background associated with the underlying event. Yet, some of the qualitative consequences
of our results are already consistent with the LHC data for di-jet asymmetry: 
the fact that the multiplicities of soft hadrons propagating at large angles are large 
and characterized by large event-by-event fluctuations
 \cite{Aad:2010bu,Chatrchyan:2011sx,Khachatryan:2015lha}.
 
Let us finally recall the main assumptions underlying our present analysis, to clarify its
limitations and suggest directions of improvement for further studies  (which however
will most likely require numerical simulations). First, the medium
has been described as a static quark-gluon plasma, characterized by a
homogeneous transport coefficient $\hat q$. In view of the phenomenology,
one must extend this set-up to an expanding medium, with a time-dependent and possibly
also space-dependent distribution for $\hat q$, that could be dynamically generated
via elastic collisions. A suitable framework in that sense is provided
by the AMY kinetic equations \cite{Arnold:2002zm} which lie
at the basis of the Monte-Carlo event generator MARTINI \cite{Schenke:2009gb}.

Second, we have limited ourselves to a leading-order formalism within
perturbative QCD at weak coupling. However, as recently understood, 
there are important quantum corrections --- notably, double-logarithmic corrections
to $\hat q$ from medium-induced radiation 
\cite{Wu:2011kc,Liou:2013qya,Iancu:2014kga,Blaizot:2014bha,
Wu:2014nca} and thermal corrections of $\order{g}$
to the kinetic equations \cite{Ghiglieri:2015ala,Ghiglieri:2015zma} 
--- that are by now available and could be used to improve our current estimates.
In particular, the double-logarithmic corrections are non-local and introduce an
additional dependence upon the medium size $L$, in the form of an `anomalous dimension'
\cite{Liou:2013qya,Iancu:2014kga,Blaizot:2014bha}.

Furthermore, we have neglected the virtuality of the leading particle and the associated
vacuum-like radiation. Whereas one does not expect the vacuum-like radiation to directly
contribute to the energy loss at large angles, it may {\em indirectly} do so, by 
producing additional sources at small angles. Besides, this is itself a random process,
which introduces additional fluctuations. We are not aware of any analytic formalism
allowing for the simultaneous treatment of the parton virtualities and the in-medium collisions, and
of their combined effect in triggering radiation. However, this becomes possible (at least, modulo
some approximations) within Monte-Carlo event generators like JEWEL  \cite{Zapp:2012ak}.
It was indeed in that Monte-Carlo context that the importance of fluctuations for the in-medium
jet evolution has been first pointed out \cite{Milhano:2015mng}. That numerical approach has also the
virtue to allow for additional sources of fluctuations which are known to be important for the phenomenology,
like those in the geometry of the hard process and the distance $L$ travelled by a jet
through the medium (a quantity that has been treated as fixed in our analysis). We hope that our
present analytic findings, although obtained in a somewhat idealized set-up, will inspire more
systematic Monte-Carlo studies and thus open the way to
realistic applications to the phenomenology.

\section*{Acknowledgments}
\vspace*{-0.3cm}
The work of M.A.E. is supported in part by the European Research Council 
under the Advanced Investigator Grant ERC-AD-267258. 

\appendix
\section{The evolution equation for the $p$-body density}
\label{app}
In this appendix we will derive Eq.~(\ref{eqNp}) for the evolution of the $p$-point correlation
function starting from Eq.~(\ref{eqZ}) for the generating functional. 
Using the definition \eqref{Np} of $\mathcal{N}^{(p)}$, we can write
\begin{eqnarray}
&&\frac{\partial}{\partial\tau}\mathcal{N}^{(p)}(x_1,x_2,\cdots,x_p|\tau)=\nonumber\\
&&\frac{\delta^p}{\delta u(x_1)\delta u(x_2)\cdots\delta u(x_p)}\,
\left\{\int\rmd z \int\rmd x\,\mcal{K}(z,x)\big[u(zx) u((1-z)x)- u(x)\big]\,
   \frac{\delta Z_\tau[u]}{\delta u(x)}\right\}_{u=1}\,.
\end{eqnarray}
The non-trivial part of the computation is to perform the $p$-th order functional derivative in the second line. To simplify the notation we define the function $T_\tau[u]$ as
\begin{equation}
T_\tau[u]\equiv \frac{\partial Z_\tau[u]}{\partial\tau}=\int\rmd z \int\rmd x\,\mcal{K}(z,x)\big[u(zx) u((1-z)x)- u(x)\big]\,
   \frac{\delta Z_\tau[u]}{\delta u(x)}\,.
\end{equation}
We need to compute the $p$-th order functional derivative of this function. As 
an example, we show the results for the first two derivatives. For $p=1$,
\begin{align}
\frac{\delta T_\tau[u]}{\delta u(x_1)}&=2\int\frac{\,\rmd z}{z}\,\mcal{K}\left(z,\frac{x_1}{z}\right)\frac{\delta Z_\tau[u]}{\delta u\left(\frac{x_1}{z}\right)}u\left(\frac{(1-z)x_1}{z}\right)-\int\,\rmd z\,\mcal{K}(z,x_1)\frac{\partial Z_\tau[u]}{\delta u(x_1)}\nonumber\\
&+\int\rmd z \int\rmd x\,\mcal{K}(z,x)\left[u(zx) u((1-z)x)- u(x)\right]\,
   \frac{\delta^2 Z_\tau[u]}{\delta u(x_1)\delta u(x)}\,,
\end{align}
where we have also used the symmetry property $\mcal{K}(1-x,z)=\mcal{K}(x,z)$. Setting $u=1$ in the previous equation one immediately obtains Eq.~(\ref{eqN}). Now we do the same for $p=2$
\begin{eqnarray}
&&\frac{\delta^2 T_\tau[u]}{\delta u(x_1)u(x_2)}=2\int\frac{\,\rmd z}{z}\mcal{K}\left(z,\frac{x_1}{z}\right)\frac{\delta^2 Z_\tau[u]}{\delta u\left(\frac{x_1}{z}\right)u(x_2)}u\left(\frac{(1-z)x_1}{z}\right)\nonumber\\
&&+2\int\frac{\,\rmd z}{z}\mcal{K}\left(z,\frac{x_2}{z}\right)\frac{\delta^2 Z_\tau[u]}{\delta u\left(\frac{x_2}{z}\right)u(x_1)}u\left(\frac{(1-z)x_2}{z}\right)\nonumber\\
&&+\frac{2}{x_1+x_2}\mcal{K}\left(\frac{x_1}{x_1+x_2},x_1+x_2\right)\frac{\partial Z_\tau[u]}{\partial u(x_1+x_2)}\nonumber\\
&&-\int\,\rmd z(\mcal{K}(z,x_1)+\mcal{K}(z,x_2))\frac{\partial^2 Z_\tau[u]}{\delta u(x_1)\delta u(x_2)}\nonumber\\
&&+\int\rmd z \int\rmd x\,\mcal{K}(z,x)\left[u(zx) u((1-z)x)- u(x)\right]\,
   \frac{\delta^3 Z_\tau[u]}{\delta u(x_1)\delta u(x_2)\delta u(x)}\,.
\end{eqnarray}
Setting $u=1$ we recover Eq.~(\ref{eqN2}). After observing the pattern that emerges 
for $p=1$ and $p=2$,
we can do an \textit{Ansatz} for the derivative of order $p$ and check that it is fulfilled:
\begin{eqnarray}
&&\frac{\delta^p T_\tau[u]}{\delta u(x_1)\cdots\delta u(x_p)}=\sum_{i=1}^p\int\,\rmd z\left[\frac{2}{z}\mathcal{K}\left(z,\frac{x_i}{z}\right)\frac{\delta^p Z_\tau[u]}{\delta u(x_1)\cdots\delta u\left(\frac{x_i}{z}\right)\cdots\delta u(x_p)}u\left(\frac{(1-z)x_i}{z}\right)\right.\nonumber \\
&&\left.-\mathcal{K}(z,x_i)\frac{\delta^p Z_\tau[u]}{\delta u(x_1)\cdots\delta u(x_p)}\right]\nonumber\\
&&2+\sum_{i=2}^p\sum_{j=1}^{i-1}\frac{1}{x_i+x_j}\mathcal{K}\left(\frac{x_i}{x_i+x_j},x_i+x_j\right)\frac{\delta^{p-1}Z_\tau[u]}{\delta u(x_1)\cdots\delta u(x_i+x_j)\cdots\delta u(x_p)}\nonumber\\
&&+\int\,\rmd z\int\,dx\mathcal{K}(z,x)[u(zx)u((1-z)x)-u(x)]\frac{\delta^{p+1}Z_\tau[u]}{\delta u(x_1)\cdots \delta u(x_p)\delta u(x)}\,.
\end{eqnarray}
The validity of this equation can be checked by induction using Eq.~(\ref{eqZ}). 
Setting $u=1$ in the above, we finally obtain the Eq.~(\ref{eqNp}). 

\section{A recursive construction for $\mathcal{N}^{(p)}$}
\label{sec:appNN}

In this appendix we shall describe the derivation of Eq.~(\ref{Npexact}),
 which is one of the main results of this paper. To that aim, it is useful to
 introduce a linear operator $I(x,\lambda,\tau-\tau')[f]$ that maps a function $f(x)$ into another function of $x$, $\lambda$ and $\tau-\tau'$ :
\begin{equation}\label{defI}
I(x,\lambda,\tau-\tau')[f]\equiv\int_x^\lambda\frac{\,\rmd \xi}{\xi^{5/2}}\mathcal{N}\left(\frac{x}{\xi},\frac{\tau-\tau'}{\sqrt{\xi}}\right)f(\lambda-\xi)\,.
\end{equation}
We shall need the action of this operator on the one-parameter family of functions
$f^\alpha(x)\equiv \frac{1}{\sqrt{x}}\rme^{-\frac{\pi\alpha^2}{x}}$. These functions 
are self-similar under the operation $I$, in the sense that 
\beq\label{falpha}
I(x,\lambda,\tau-\tau')[f^\alpha]=\frac{1}{x^{3/2}}f^{\tau-\tau'+\alpha}(\lambda-x)\,.\eeq
This can be checked as follows: one has
\begin{equation}
\int_x^\lambda\frac{\,\rmd \xi}{\xi^{5/2}}\,
\mathcal{N}\left(\frac{x}{\xi},\frac{\tau-\tau'}{\sqrt{\xi}}\right)\frac{\rme^{-\frac{\pi\alpha^2}{\lambda-\xi}}}{\sqrt{\lambda-\xi}}=\frac{\tau-\tau'}{x^{3/2}}\int_x^\lambda\frac{\,\rmd \xi}{(\xi-x)^{3/2}}\frac{1}{\sqrt{\lambda-\xi}}\rme^{-\frac{\pi(\tau-\tau')^2}{\xi-x}}\rme^{-\frac{\pi\alpha^2}{\lambda-\xi}}\,.
\end{equation}
The integral in the r.h.s. can be simplified with the change of variables $u=\frac{\xi-x}{\lambda-\xi}$,
which gives
\begin{equation}
\frac{\tau-\tau'}{x^{3/2}(\lambda-x)}\rme^{-\frac{\pi[(\tau-\tau')^2+\alpha^2]}{\lambda-x}}\int_0^\infty\frac{\,du}{u^{3/2}}\rme^{-\frac{\pi(\tau-\tau')^2}{(\lambda-x)u}-\frac{\pi\alpha^2 u}{\lambda-x}}\,,
\end{equation}
or, after also using Eq.~(B.6) of \cite{Escobedo:2016jbm},
\begin{equation}
\frac{1}{x^{3/2}\sqrt{\lambda-x}}\rme^{-\frac{\pi(\tau-\tau'+\alpha)^2}{\lambda-x}}\,,
\end{equation}
which is the result that we anticipated. 

\subsection{The case $p=2$}
\label{sec:appN}
It is useful to observe that the source term in Eq.~(\ref{N2sol}) can be written as
\begin{equation}
S^{(2)}(\xi_1,\xi_2|\tau')=-\frac{1}{2\pi \xi_1^{3/2}\xi_2^{3/2}}\frac{\rmd}{\rmd \tau'}f^{\tau'}(1-\xi_1-\xi_2)\,,
\end{equation}
with the function $f^\tau$ as introduced above \eqn{falpha}. By using  \eqn{defI}, it is easy to see that Eq.~(\ref{N2sol}) can be viewed as the result of applying twice the operator $I$ on
$f^\tau$ and then integrating over $\tau'$ :
\begin{align}
\mathcal{N}^{(2)}(x_1,x_2|\tau)&=-\frac{1}{2\pi}\int_0^\tau\,\rmd \tau'\lim_{\tau_2\to\tau'}\frac{\rmd}{\rmd \tau_2}\int_{x_1}^1\frac{\,\rmd \xi}{\xi^{5/2}}\mathcal{N}\left(\frac{x_1}{\xi_1},\frac{\tau-\tau'}{\sqrt{\xi_1}}\right)I(x_2,1-\xi_1,\tau-\tau')[f^{\tau_2}]\nonumber\\
&=-\frac{1}{2\pi x_2^{3/2}}\int_0^\tau\,\rmd \tau'\lim_{\tau_2\to\tau'}\frac{\rmd}{\rmd \tau_2}I(x_1,1-x_2,\tau-\tau')[f^{\tau-\tau'+\tau_2}]\nonumber\\
&=\frac{1}{2\pi(x_1x_2)^{3/2}}\,
\frac{\rmd}{\rmd \tau'}\int_0^\tau\,\rmd \tau'f^{2\tau-\tau'}(1-x_1-x_2)\,,
\label{eq:N2}
\end{align}
where the action of the operator $I$ has been (twice) computed according to  \eqn{falpha}. 
From this result it is straightforward to obtain Eq.~(\ref{N2exact}).

\subsection{The general case $p\ge 2$}
Inspired by the above result for $p=2$,
we make the hypothesis, which will turn out to be true, that the multi-gluon density
admits the following general structure for any $p\ge 1$ :
\begin{equation}
\mathcal{N}^{(p)}(x_1,\cdots,x_p|\tau)=\frac{1}{(x_1\cdots x_p)^{3/2}}d^{(p)}(1-\sum_{i=1}^px_i|\tau)\,.
\label{eq:Ndecomp}
\end{equation}
Under this assumption, we can write the source term in  Eq.~(\ref{Npsol})  as 
\begin{equation}\label{ansatzSp}
S^{(p)}(x_1,\cdots,x_p|\tau)=\left(\begin{array}{c}
p \\
2\end{array}\right)\frac{1}{(x_1\cdots x_p)^{3/2}}\,d^{(p-1)}(1-\sum_{i}^p x_i|\tau)\,.
\end{equation}
The appearance of the combinatorial number can be physically understood as a consequence of the symmetry of the problem when interchanging the labels $i$ and $j$ of two gluons. In view of this, one can understand the right-hand side of Eq.~(\ref{Npsol}) as the result of acting $p$ times on $d^{(p-1)}$ with the operator $I$ (one action for each coordinate $\xi_i$) 
and then integrating over $\tau'$. This implies that if $d^{(p-1)}$ can be written in terms of the family of functions $f^\alpha$, then the same is true
for $d^{(p)}$. We will now perform the computation of $\mathcal{N}^{(3)}$ taking into account this structure. To that aim, we first read the expression of $d^{(2)}(l|\tau)$ from the last line in \eqn{eq:N2},
namely
\begin{equation}
d^{(2)}(l|\tau)=\frac{1}{2\pi}\int_0^\tau\,\rmd \tau_1\frac{\rmd}{\rmd \tau_1}f^{2(\tau-\tau_1)+\tau_1}(l)\,.
\end{equation}
By inserting this result into \eqn{ansatzSp} with $p=3$, one deduces
\begin{equation}
S^{(3)}(x_1,x_2,x_3|\tau)=\frac{3}{2\pi(x_1 x_2 x_3)^{3/2}}\big[f^\tau(1-x_1-x_2-x_3)-f^{2\tau}(1-x_1-x_2-x_3)\big]\,.
\end{equation}
Doing a computation analogous to that in Eq.~(\ref{eq:N2}), we obtain
\begin{equation}
\mathcal{N}^{(3)}(x_1,x_2,x_3|\tau)=\frac{3}{2\pi(x_1x_2x_3)^{3/2}}\int_0^\tau\,\rmd \tau'
\big[f^{3(\tau-\tau')+\tau'}(1-x_1-x_2-x_3)-f^{3(\tau-\tau')+2\tau'}(1-x_1-x_2-x_3)\big]\,,
\end{equation}
which confirms that $\mathcal{N}^{(3)}$ can be written in the form of Eq.~(\ref{eq:Ndecomp}) with
\begin{equation}
d^{(3)}(l|\tau)=\frac{3}{2\pi}\int_0^\tau\,\rmd \tau_2\int_0^{\tau_2}\,\rmd \tau_1\frac{\rmd}{\rmd \tau_1}f^{3(\tau-\tau_2)+2(\tau_2-\tau_1)+\tau_1}(l)\,.
\end{equation}
Based on that, it is easy to guess that 
\begin{equation}
d^{(p)}(l|\tau)=A_p\int_0^\tau\,\rmd \tau_{p-1}\cdots\int_0^{\tau_2}\,\rmd \tau_1\frac{\rmd}{\rmd \tau_1}f^{p\tau-\sum_{i=1}^{p-1}\tau_i}(l)\,,
\label{eq:gendp}
\end{equation}
where $A_p$ some proportionality constant that only depends on $p$. We can check explicitly that if this assumption is fulfilled for $p-1$ then it is also fulfilled for $p$ with $A_p=\left(\begin{array}{c} p\\2\end{array}\right)A_{p-1}$. Using this together with $A_2={1}/{2\pi}$, we find
\begin{equation}
A_p=\frac{(p!)^2}{2^p\pi p}\,.
\label{eq:Ap}
\end{equation}
Combining eqs. (\ref{eq:Ndecomp}), (\ref{eq:gendp}) and (\ref{eq:Ap}) we can recover Eq.~(\ref{Npexact}) after performing the change of variables $\frac{\tau_i}{\sqrt{1-\sum_{j=1}^p x_j}}\to l_i$.

\section{Computation of $h_p(l)$ in two limiting cases}

The equation (\ref{hp}) can be rewritten in the following way
\begin{equation}
h_p(l)=\int_0^l\,\rmd l_{p-1}\cdots\int_0^{l_2}\,\rmd l_1 f(l|l_1,\cdots,l_{p-1})\,,
\end{equation}
where $f$ has the property that $f(l|l_1,\cdots,l_i,\cdots,l_j,\cdots,l_{p-1})=f(l|l_1,\cdots,l_j,\cdots,l_i,\cdots,l_{p-1})$ for any $i$ and $j$. Therefore we can rewrite the expression for $h_p$ as 
\begin{equation}
h_p(\ell)=\frac{1}{(p-1)!}\int_0^\ell\,\rmd\ell_{p-1}\cdots\int_0^{\ell}\,\rmd\ell_1\Big(p\ell-\sum_{i=1}^{p-1}\ell_i\Big)\,
\rme^{-\pi\big(p\ell-\sum_{j=1}^{p-1}\ell_j\big)^2}\,.
\end{equation}
We can also apply the change of variables $l_i\to l\lambda_i$ to obtain
\begin{equation}
h_p(l)=\frac{l^p}{(p-1)!}\int_0^1\,\rmd \lambda_{p-1}\cdots\int_0^1\,\rmd \lambda_1\left(p-\sum_{i=1}^{p-1}\lambda_i\right)\rme^{-\pi l^2\big(p-\sum_{j=1}^{p-1}\lambda_j)^2}\,.
\label{eq:hpsym}
\end{equation}
Using this equation, we shall now study the two limiting cases of physical interest.

\subsection{$h_p(l)$ in the limit $\pi (pl)^2\ll 1$}
\label{sec:smalll}
The equation (\ref{eq:hpsym}) can be simplified if one can substitute the exponential by $1$ for all the values of $\lambda_i$ inside the domain. This will precisely happen when $\pi (pl)^2\ll 1$ as this is the maximum value of the argument of the exponential, in this case we can use that
\begin{equation}
\int_0^1\,\rmd \lambda_{p-1}\cdots\int_0^1\,\rmd \lambda_1\left(p-\sum_{i=1}^{p-1}\lambda_i\right)=\int_0^1\,\rmd \lambda_{p-2}\cdots\int_0^1\,\rmd \lambda_1\left(p-\sum_{i=1}^{p-2}\lambda_i\right)-\frac{1}{2}\,.
\end{equation}
Iterating this formula $p-1$ times, we obtain
\begin{equation}
\int_0^1\,\rmd \lambda_{p-1}\cdots\int_0^1\,\rmd \lambda_1\left(p-\sum_{i=1}^{p-1}\lambda_i\right)=\frac{p+1}{2}\,,
\end{equation}
which leads to
\begin{equation}
h_p(l)\simeq\frac{l^p (p+1)}{2(p-1)!}\,.
\end{equation}

\subsection{$h_p(l)$ in the limit $\pi l^2\gg 1$}
\label{sec:largel}
In this case it is convenient to perform the change of variables $\lambda_i\to 1-y_i$
\begin{equation}
h_p(l)=\frac{l^p}{(p-1)!}\int_0^1\,\rmd y_{p-1}\cdots\int_0^1\,\rmd y_1\left(1+\sum_{i=1}^{p-1}y_i\right)\rme^{-\pi l^2\big(1+\sum_{j=1}^{p-1}y_j\big)^2}\,.
\end{equation}
When $\pi l^2\gg 1$, the integral will be dominated by the region $y_i\sim\frac{1}{\pi l^2}\ll 1$, therefore we can expand for small values of $y_i$
\begin{equation}
h_p(l)\simeq\frac{l^p}{(p-1)!}\,\rme^{-\pi l^2}\left(\int_0^1\,\rmd y\,\rme^{-2\pi l^2 y}\right)^{p-1}\,.
\end{equation}
Given that the integral is dominated by small values of $y$, 
we can change the integration region from $[0,1]\to[0,\infty)$ 
introducing a negligible error; therefore we can write
\begin{equation}
h_p(l)\simeq\frac{l^p}{(p-1)!}\,\rme^{-\pi l^2}\left(\frac{1}{2\pi l^2}\right)^{p-1}=\frac{\rme^{-\pi l^2}}{(p-1)!l^{p-2}(2\pi)^{p-1}}\,.
\end{equation}
\bigskip

\providecommand{\href}[2]{#2}\begingroup\raggedright\endgroup

\end{document}